\documentstyle[12pt,graphicx]{article}

\begin{document}

\title{\bf The rational parts of one-loop QCD  amplitudes I: The general
formalism}

\author{Zhi-Guang Xiao\thanks{E-mail: zhgxiao@itp.ac.cn} $^{1,2}$,
Gang Yang\thanks{E-mail: yangg@itp.ac.cn} $^{2}$and Chuan-Jie
Zhu\thanks{E-mail: zhucj@itp.ac.cn} $^{2,3}$ }

\maketitle

\medskip\centerline{$^1$The Interdisciplinary Center of Theoretical
Studies, Chinese} \centerline{Academy of Sciences, P. O. Box 2735,
Beijing 100080, P.~R.~China}
\medskip
\centerline{$^2$Institute of Theoretical Physics, Chinese Academy
of Sciences} \centerline{P. O. Box 2735, Beijing 100080, P. R.
China}
\medskip
\centerline{$^3$Center of Mathematical Science, Zhejiang
University} \centerline{Hangzhou 310027, P. R. China}

\begin{abstract}

A general formalism for computing only the rational parts of
one-loop QCD amplitudes is developed. Starting from the Feynman
integral representation of the one-loop amplitude, we use tensor
reduction and recursive relations to compute the rational parts
directly. Explicit formulas for the rational parts are given for
all bubble and triangle integrals. Formulas are also given for box
integrals up to two-mass-hard boxes which are  the needed
ingredients to compute up to 6-gluon QCD amplitudes. We use this
method to compute explicitly the rational parts of the 5- and
6-gluon QCD amplitudes in two accompanying papers.

\end{abstract}
\newpage

\section{Introduction}
The forthcoming experimental program at CERN's Large Hadron
Collider (LHC) requires many computations at the
next-to-leading order (NLO) or one-loop with many particles as final
states \cite{Salam}. However the analytic computation of the
matrix elements is very difficult. Only for special helicity configurations
\cite{BCDK,BDDK,Mahlon} or special models \cite{BinothYukawa},
some analytic results are known for higher point amplitudes.  The
current state of the art in NLO computation is 5-point for QCD
processes and 6-point for electroweak processes
\cite{FivePoint,BernKosower,Denner}. The recent development in
tackling the multi-leg amplitudes by semi-numerical/analytic
methods shows promise for improving traditional capabilities
\cite{RKEllis,Glover,BinothA,PittauB, Passarino, Soper,
Anastasiou, Rest}. All helicity configurations for the 6-gluon
amplitude are evaluated for a single-space point \cite{RKEllis}.
These results are used to check the available analytic results
\cite{BDKA}.

Following Witten's twistor string theory \cite{Witten}, the CSW
approach \cite{CSW} and the use of maximally-helicity-violating
(MHV)\footnote{The 2 dimensional origin of the MHV amplitudes in
gauge theory was first given in \cite{Nair}} vertices
\cite{ParkeA,ParkeR}, there has been spectacular progress
\cite{Zhu,Khoze,WuZhu,BernSome,CSWA,BST,Cachazo,Rozali,BBST,BCFW,
Dunbar, BDKBoot, BernA, BernB, Luo, BoFengA, BoFengC, BoFengSix,
CachazoP, AMST, SuWu, Abe, CSWB, FordeKosower, Scratch} in the
perturbative QCD computations in the last two years or so, by
using the unitarity cut method of Bern, Dunbar, Dixon and Kosower
\cite{BDDK,BDKB,BDK} and the spinor-helicity formalism
\cite{Berends,ChineseMagic} (see \cite{Dixon} for a review). In
particular, Bedford, Brandhuber,  Spence and  Travaglini
\cite{BST,BBST} applied the MHV vertices to one-loop
calculations. Britto, Buchbinder, Cachazo, Feng and Mastrolia
\cite{BoFengA, BoFengC, BoFengSix} developed an efficient technique
for evaluating the rational coefficients in an expansion of the
one-loop amplitude in terms of scalar box, triangle and bubble
integrals (the cut-constructible part, see below and Sect.~4). By
using their technique, it is much easier to calculate the
coefficients of box integrals without doing  any integration.
Recently, Britto, Feng and Mastrolia completed the computation of
the cut-constructible terms for all the 6-gluon helicity
amplitudes \cite{BoFengSix}.

In order to complete the QCD calculation for the 6-gluon
amplitude,  the remaining challenge is to compute the rational
part of the amplitude with scalars circulating in the loop,
commonly called the $N=0$ case in a supersymmetric decomposition
of QCD amplitudes:
\begin{equation}
A^{QCD} = A^{N=4} - 4 A^{N=1~{\rm chiral}} + A^{N=0~{\rm or~
scalar}} .
\end{equation}
The above strategy of splitting the computation of the QCD amplitude
into various supersymmetric parts plus a scalar part is quite
fruitful. By using a theorem of Bern, Dunbar, Dixon and Kosower
\cite{BDK}, the supersymmetric parts are cut-constructible,
meaning that these amplitudes can be determined completely by
using 4-dimensional unitarity. Even for the scalar part which is
not cut-constructible, we can still split it into two parts: a
cut-constructible part and a rational part. As we said before, the
recent development inspired by twistor string theory has lead to
very efficient techniques to compute the cut-constructible part
\cite{BoFengA, BoFengC, BoFengSix}. To complete the program it is
quite important to have efficient and powerful methods to compute
the rational part.

There are various ways to compute the rational part. The first
approach \cite{BDKB} is to use the factorization properties by trial
and error. This is a quite effective method if the final result is
simple enough. The difficulty with this approach is that we do not
know how to automate the method to make effective use of the
advances in computer industry. The correctness of the obtained
result is almost guaranteed by checking factorizations for all
channels. For higher point amplitudes, the complexity of the
analytic results makes this method impractical.

The second approach uses the unitarity relation. In principle the
rational part can be constructed by using the $D$-dimensional
unitarity method \cite{BernMorgan,BDK,BDKC,AMST}. The problem with
this approach is that too much information is kept and tree
amplitudes in $D$-dimension are even more difficult
\cite{DixonRecent}. In fact this approach loses the simplicity of
4-dimensional helicity amplitudes as given by the MHV formula.

The third approach is the bootstrap recursive approach of Bern,
Dixon and Kosower \cite{BDKBoot}. This  approach is quite
promising and powerful. It is a streamlined approach of the first
one by adding the insights of the recent tree-level recursive method
of Britto, Cachazo, Feng and Witten \cite{BCFW}. This approach has
already produced a wealth of general results for special helicity
configurations, notably the ``split-helicity" configurations
\cite{BDKBoot, FordeKosower,BDKA}. It can also be used to compute
one-loop QCD amplitudes with general helicities as outlined in
\cite{BDKA}.

Given the complexity of the results for the cut-constructible part
of the 6-gluon amplitude \cite{BoFengSix} and its important
applications to LHC related experiments, it is quite worthy and
even mandatory to have other methods to compute the rational part
of the QCD amplitude. In particular one would like to bypass the
need of using the cut-constructible part and have an independent
method to compute the rational part. Of course, the testing ground
for any method is a complete computation of the 6-gluon QCD
amplitude where only partial results for some helicity
configurations exist. The present status for the marching  to one
loop 6-gluon QCD amplitude were summarized in
\cite{DixonRecent,BernRecentA}. For more recent developments, we
refer the reader to \cite{BernRecentB,KosowerRecent}.

In this paper we will study the problem of computing the rational
parts of one-loop amplitudes directly from Feynman integral
representations. These integrals can be written down directly by
drawing all Feynman diagrams and by using the Feynman rules. With
the present technology these can be done quite effectively by
using the various packages like GRACE \cite{GRACEFeynman},
FeynArts \cite{FeynArts} and Qgraf \cite{Qgraf} et. al. (See
\cite{Steinhauser,GRACEFeynman} for reviews.) Fortunately these
powerful methods are not needed to compute up to the 6-gluon
amplitude. For computing higher point amplitudes or 6-parton
amplitudes they may be a necessity.

It is easy to imagine that the rational part is already contained
in the integral representation of the amplitude. If one could
obtain the complete rational coefficients by doing tensor
reduction to scalar box, triangle and bubble integrals, one can
simply get the rational part by making an expansion with the
dimension $D$ around 4 ($D=4-2\epsilon$ is the parameter of
dimensional regularization). This is extraordinarily difficult
because of the complexity of tensor reduction for $N\ge5$. However
if one only needs to compute the rational part, it is not
necessary to know the complete coefficients from tensor
reductions. By the BDDK theorem \cite{BDDK}, we know that many
terms simply do not contribute to the rational part. Following
this path of thought, the remaining problem is: is there an
efficient way to compute these rational parts in one-loop QCD
amplitudes directly from Feynman integrals?

In this paper we show that there is actually a quite efficient and
powerful method to compute the rational part directly from Feynman
integrals. Because we are concerned only with the rational part of
the amplitude, there is  no need for tensor reduction all the way
down to scalar integrals. We only need tensor reduction to reduce
the degree of the numerator by 2. So the original complexity of
tensor reduction is bypassed in the computation of the rational part.

In our approach of computing the rational part, we will exploit
the theorem of Bern, Dunbar, Dixon and Kosower \cite{BDDK} and
directly extract the rational part from the one-loop Feynman
integrals. We also use the simple tensor reduction by using
spinors as developed in \cite{BDKB,PittauA,Weinzierl1}. We point
out that the tensor reduction formulas used in our computations
are actually quite simple, as one can see from
eqs.~(\ref{eqreduction}) and (\ref{eqreductiona}) in Sect.~3.

As we will demonstrate in this paper, the computation of the
rational part is reduced to tree-level like calculations. As our
method also applies to massive theory and theories with fermions,
we envisage wider applications of our method in the
computation of one-loop amplitudes, in combination with the $D=4$
unitarity method \cite{BDDK, BDKB} and the efficient technique for
computing generic unitarity cuts \cite{BoFengA}. The once most
difficult part of the one-loop amplitude can actually be attacked
by the traditional technique.

In this paper and the accompanying two papers \cite{xyzii,xyziii},
we will develop our method and apply it to the computation of the
rational parts of the one-loop 5- and 6-gluon QCD amplitudes. This
paper mainly deals with the general theoretical formalism of the
method. In \cite{xyzii}, we show the efficiency of the method by
computing the rational parts for the 5-gluon amplitudes for the two
MHV helicity configurations by giving most of the intermediate
steps. In \cite{xyziii}, we will present the results for the
rational parts of the 6-gluon amplitudes for the two MHV and two
NMHV helicity configurations. The rational parts of the 6-gluon
amplitudes for the ``split helicity" configurations are known
already \cite{BDKBoot,BDKA}. Recently all one-loop maximally
helicity violating gluonic amplitudes were computed by Berger,
Bern, Dixon, Forde and Kosower \cite{BDKE}. We refer the reader to
\cite{BDKE,xyziii} for details about the explicit analytic results
and comparisons.

This paper is organized as follows: in Sect. 2 we set up our notation
for spinor products and composite currents for sewing trees to
the loop. Some simple tensor reduction formulas are given in Sect.
3. Starting from Sect. 4, we begin to develop the method of
extracting the rational parts of Feynman integrals. We use the
recursive approach to compute any Feynman integral as developed in
\cite{BDKReduction}. In Sect. 6 and 7 we give explicit results for
triangle and box integrals. In Sect.~8 we compute the correction
terms to the naive $D=4$ tensor reduction of box and
triangle integrals, which arise from the ultra-violet divergent
part of the box and triangle amplitudes.

\section{Notation}
We mainly follow the notation of BDK \cite{BDKBoot} and the
QCD-literature convention for the square bracket $[i\,j]$. By
abusing of notation the product between 2 holomorphic spinors or 2
anti-holomorphic spinors is formed by a round bracket:
\begin{equation}
(\lambda_i, \lambda_j) = \langle i \, j \rangle,  \qquad \qquad
(\tilde\lambda_i, \tilde\lambda_j) =  [i \, j ] .
\end{equation}
The scalar product between 2 vectors (written either in 4d vector
notation or in 2 spinor notation) is also denoted by a round
bracket. We have
\begin{eqnarray}
(\lambda_{i_1}\tilde\lambda_{j_1},\lambda_{i_2}\tilde\lambda_{j_2} ) & = &
(\lambda_{i_1}, \lambda_{i_2})\, (\tilde\lambda_{j_2},\tilde\lambda_{j_1})  =
\langle i_1\,i_2\rangle [j_2\, j_1] ,
\\
2 \, k_i \cdot k_j  & = &(\lambda_{i} \tilde\lambda_i, \lambda_{j} \tilde\lambda_j)
= \langle i\, j \rangle\, [j\, i] .
\end{eqnarray}
For spinor strings, we simply use $\langle \lambda_i|(k_a+k_b)|\lambda_j
\rangle$ or $\langle i|(k_a+k_b)|j]$ to denote $\langle
i^-|(a+b)|j^-\rangle$:
\begin{eqnarray}
\langle \lambda_i|(k_a+k_b)|\tilde\lambda_j \rangle & = &  \langle i|(k_a+k_b)|j]
 = \langle i|( a+ b)|j]
\nonumber \\
& = & \langle i^-|(a+b)|j^-\rangle = \langle i
\,a\rangle \, [a\,j] + \langle i\,b\rangle \, [b\,j] .
\end{eqnarray}
We do not use gamma matrix traces. Instead we use bra and ket
notation  with multiple insertions of momenta:
\begin{equation}
\langle i|k_1\, k_2 \cdots k_n|j\rangle = \left\{
\begin{array}{ll}
 \langle i\, 1\rangle [1\, 2] \cdots [n\, j], & n = {\rm odd}, \cr
 \langle i\, 1\rangle [1\, 2] \cdots \langle n\, j\rangle, & n = {\rm even}.
\end{array}
\right.
\end{equation}
Sometimes we also write $\langle i |K | j \rangle $ for $\langle i
|K | j ] $, with the understanding that sometimes the last $j$
should actually stand for $\tilde\lambda_j$ and with the bracket
$]$. For example we have
\begin{equation}
\langle i|k_1\, k_2\, k_3|j] = \langle i\, 1\rangle \,
[1\, 2] \, \langle 2 \, 3\rangle \, [3 \, j] .
\end{equation}
For $i=j$ the above notation is just the gamma matrix trace.
For simplicity we will not write the slash:
\begin{equation}
\langle i|k_1\, k_2\, k_3|i] = {\rm tr}_-( k_i\, k_1\, k_2\, k_3)
.
\end{equation}
We note that the above notation only happens for an odd number of
momenta inserted between 2 spinors (one holomorphic and one
anti-holomorphic). Of course the momentum can be either massless
or a sum of several massless momenta.

The sums of cyclicly consecutive external momenta are denoted
generically  by $K_i$ in a Feynman diagram. In our explicit
computation we use $k_{12}=k_1+k_2$ and $k_{234}=k_2+k_3+k_4$,
etc. The kinematic variables are denoted as $s_{12}=(k_1+k_2)^2$
and $s_{123}=(k_1+k_2+k_3)^2$ in a self explaining notation. For
6-gluon case we also have  $s_{123}=(k_4+k_5+k_6)^2$ by momentum
conservation.

\begin{figure}[ht]
\centerline{\includegraphics[height=2.5cm]{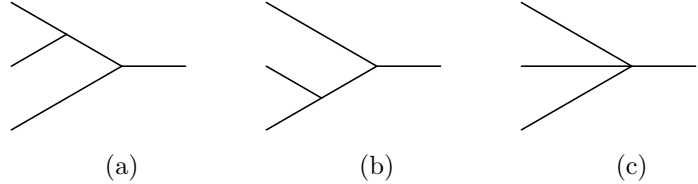} } \caption{The
composition of 3 external particles in tree amplitudes.}
\label{One}
\end{figure}

\begin{figure}[ht]
\centerline{\includegraphics[height=4.5cm]{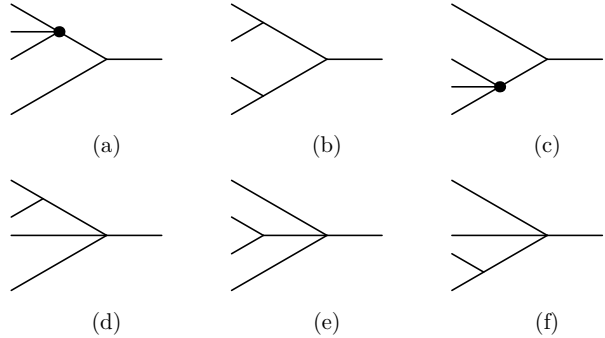} } \caption{The
composition of 4 external particles in tree amplitudes. The blob
denotes an expansion as given in Fig.~\ref{One}. The explicit
expression of $\epsilon_{i (i+1)(i+2)(i+3)}$ will not be given.}
\label{Two}
\end{figure}

For sewing trees to the loop, we define the following composite
currents or polarization vectors:
\begin{eqnarray}
\epsilon_{i (i+1)} & = & P(\epsilon_i,k_i; \epsilon_{i+1},k_{i+1})
\equiv
  {1\over (k_i+k_{i+1})^2 } \Big(
(\epsilon_{i}, k_{i+1})\, \epsilon_{i+1} \nonumber \\
&-&  (\epsilon_{i+1}, k_{i})\, \epsilon_{i} +
  {1\over 2} (\epsilon_i,\epsilon_{i+1}) \,
(k_i - k_{i+1}) \Big) , \\
\epsilon_{i(i+1)(i+2)} & = &   P(\epsilon_{i(i+1)},k_{i(i+1)};
\epsilon_{i+2},k_{i+2}) +  P(\epsilon_i
,k_{i};\epsilon_{(i+1)(i+2)},k_{(i+1)(i+2)})
\nonumber \\
&  & \hskip -2cm + {1\over s_{i(i+1)(i+2)}  }\, \left(
(\epsilon_i,\epsilon_{i+2}) \, \epsilon_{i+1} - {1\over2}\,
(\epsilon_i,\epsilon_{i+1}) \, \epsilon_{i+2} - {1\over2}\,
(\epsilon_{i+1},\epsilon_{i+2}) \, \epsilon_{i } \right)  ,
\end{eqnarray}
where $s_{i(i+1)(i+2)} = (k_i+k_{i+1} + k_{i+2})^2$. The above
procedure is a simplified version of the general recursive
calculation of the tree-level $n$-gluon amplitudes
\cite{BerendsGiele}. We note that $\epsilon_{i(i+1)}$ is
anti-symmetric and $\epsilon_{i(i+1)(i+2)}$ is symmetric under the
reversing of the order of the particles. This generalizes to
composite currents with more legs, which we have not written down explicitly. The
diagrammatic representations of $\epsilon_{i(i+1)(i+2)}$ and
$\epsilon_{i(i+1)(i+2)(i+3)}$ are given in Figs.~\ref{One} and
\ref{Two}.

In the formalism of \cite{ChineseMagic} (see \cite{Dixon,ParkeR} for reviews), the gluon polarization vectors are defined as
\begin{equation}
\varepsilon_\mu^{(+)}(k;q)={\langle q^-|\gamma_\mu|k^-\rangle\over \sqrt{2}\langle q^-|k^+\rangle}\,,
\hspace{1cm}
\varepsilon_\mu^{(-)}(k;q)={\langle q^+|\gamma_\mu|k^+\rangle\over \sqrt{2}\langle k^+|q^-\rangle}\,,
\end{equation}
where $k$ is the momentum of the polarization null vector
and $q$ is the reference null vector.
In terms of the holomorphic and anti-holomorphic spinors
( $k=\lambda \tilde{\lambda}$ and $q=\eta\tilde{\eta}$), these polarization vectors
can be recast as
\begin{equation}
\varepsilon_{\alpha\dot{\beta}}^{(+)}(k;q)={\sqrt{2}\,\eta_\alpha\tilde{\lambda}_{\dot{\beta}}\over \langle\eta \lambda\rangle}\,,
\hspace{1cm}
\varepsilon_{\alpha\dot{\beta}}^{(-)}(k;q)={\sqrt{2}\,\lambda_\alpha\tilde{\eta}_{\dot{\beta}}\over [\lambda\eta ]}\,.
\end{equation}
In our notation, we will use $\epsilon$ to denote the polarization vectors,
and the relations between $\epsilon$ and $\varepsilon$ is $\varepsilon=\sqrt{2}\,\epsilon$.
The troublesome $\sqrt{2}$ will be absorbed in the overall coefficient of the amplitude.
Given this notation, the bracket product of polarizations is more natural than the dot product :
\begin{equation}
(\epsilon_j^-(k_j;q_j),\epsilon_l^-(k_l,q_l))=\varepsilon_j^-(k_j;q_j)\cdot\varepsilon_l^-(k_l,q_l)
={\langle j\,l\rangle[q_lq_j]\over [j\,q_j][l\,q_l]}\,.
\end{equation}
In our calculations, it is also convenient to omit the denominators in the definition
of the polarizations, and reinstate them in the last step.

\section{Tensor reduction of the one-loop amplitudes}

There is a vast literature on this subject. The original
Passrino-Veltman approach \cite{PassarinoVeltman} is quite general
but it is not quite practical to obtain compact analytic results.
In fact the tensor reduction relations we will use for our
calculations of the 5- and 6-gluon amplitudes are quite simple. It
is based on the BDK trick \cite{BDKB} of multiplying and dividing
by spinor square roots. To make more effective use of this trick,
we have purposely chosen the reference momenta to make the tensor
reduction simple. See \cite{xyziii} for further details about the
specific choices of the reference momenta and the tensor reductions
involved in the computation of (the rational parts of) the 6-gluon
amplitudes.

\begin{figure}[ht]
\centerline{\includegraphics[height=2cm]{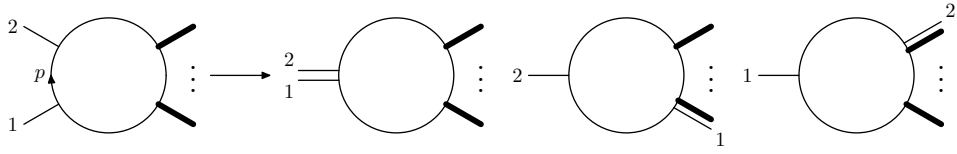} }
\caption{The tensor reduction for two adjacent same helicities
generally gives three terms.} \label{Three}
\end{figure}

There are basically only two different cases to consider. As shown
in Fig.~\ref{Three}, the two polarization vectors have the same
helicity. If we choose the same reference momentum (denoted by
the spinor $\eta$), we have
\begin{eqnarray}
(\eta\tilde\lambda_1, p)\, (\eta\tilde\lambda_2, p) & = & - {
(\eta\tilde\lambda_{k_{12}}^{(\eta)}, p + k_1) \over \langle 1\, 2
\rangle }
\, I^{(2)} \nonumber \\
& + & {\langle\eta\, 1\rangle \over \langle 1\, 2 \rangle } \,
(\eta\tilde\lambda_1, p) \, I^{(3)} + {\langle\eta\, 2\rangle
\over \langle 1\, 2 \rangle } \, (\eta\tilde\lambda_2, p) \,
I^{(1)} , \label{eqreduction} \\
\tilde\lambda_{k_{12}}^{(\eta)} & = & \langle\eta\, 1\rangle
\tilde\lambda_1 + \langle\eta\, 2\rangle \tilde\lambda_2,
\end{eqnarray}
where $I^{(1)} =(p+k_1)^2$, $I^{(2)} = p^2$ and
$I^{(3)}=(p-k_2)^2$ are various inverse propagators. The above
tensor reduction formula is shown diagrammatically in
Fig.~\ref{Three}, omitting the relevant factors.

In deriving eq.~(\ref{eqreduction}), we assumed that $p$ is a  four-dimensional vector.
Because pentagon and higher point one-loop amplitudes are ultra-violet convergent,
the use of the above formula in tensor reduction is correct in dimensional
regularization up to infinitesimal terms. However one must be quite careful to apply the
above formula to the tensor reduction of the box and triangle tensor integrals because the
difference is a finite rational part. Some correction terms must be included for
tensor reduction with box and triangle tensor integrals. This also applies to the following
tensor reduction formulas given later in this section.
We will compute these correction terms later in Sect.~8.

\begin{figure}[ht]
\centerline{\includegraphics[height=4cm]{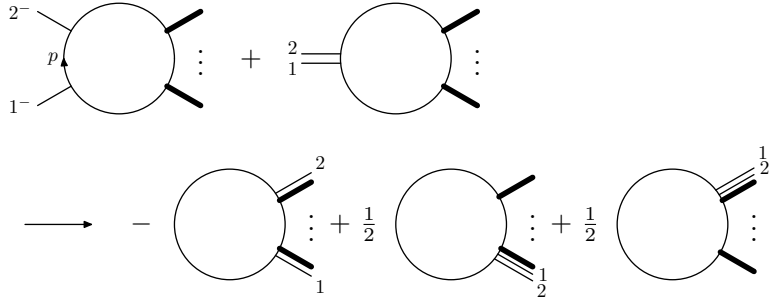} }
\caption{For two adjacent same helicities, the tensor reduction for
the combination of two diagrams is even simpler by a judicious
choice of the reference momenta.} \label{Four}
\end{figure}

An even simpler version of the above tensor reduction relation is
to consider a combination of two diagrams together as shown in
Fig.~\ref{Four}. The reduction formula is:
\begin{eqnarray}
 {(\epsilon_1,p+k_1)(\epsilon_2,p ) \over (p+k_1)^2
 p^2(p-k_{2 })^2} &  + & { (\epsilon_{12}, p+k_1) -
(\epsilon_1,\epsilon_2)/2 \over (p+k_1)^2 (p-k_{2})^2}
\nonumber \\
& = & - {1\over p^2} + { {1/ 2}\over  (p+k_1)^2} + {{1/2}\over
(p-k_{2})^2} , \label{eqreductiona}
\end{eqnarray}
for $\epsilon_1  = \lambda_1\tilde\lambda_2$ and $\epsilon_2=
\lambda_2\tilde\lambda_1$. All the factors appearing on the
left-hand side of the above equations are read off directly from
Feynman rules. These tensor reduction relations show quite clearly
the simplicity of the diagrams when there are adjacent particles
with the same helicity.

\begin{figure}[ht]
\centerline{\includegraphics[height=3cm]{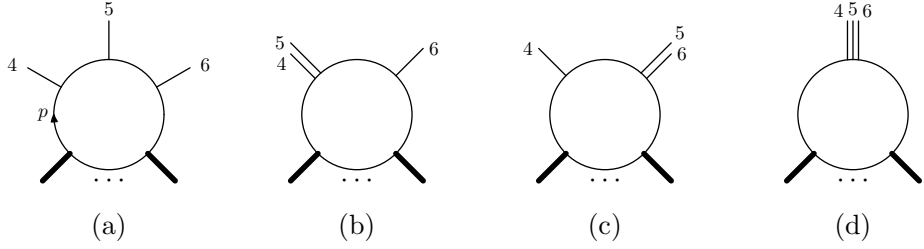} } \caption{For
three adjacent same helicities, the tensor reduction for the
combination of these four diagrams is also quite simple if we
choose the reference momenta appropriately.} \label{Five}
\end{figure}

For three adjacent particles with the same helicity gluons, we
choose the following polarization vectors (omitting an overall
factor for each polarization vector):
\begin{equation}
\epsilon_4 = \lambda_5\tilde\lambda_4, \qquad \epsilon_5 =
\eta\tilde\lambda_5, \qquad \epsilon_6 = \lambda_5\tilde\lambda_6.
\end{equation}
Then we have
\begin{eqnarray}
\epsilon_{45} & = &  - \eta\tilde\lambda_4 + \frac{1}{2}\,
\frac{\langle\eta\, 5\rangle}{\langle4\, 5\rangle}\, k_{45},
\\
\epsilon_{56} & = &  \eta\tilde\lambda_6 - \frac{1}{2}\,
\frac{\langle\eta\, 5\rangle}{\langle6\, 5\rangle}\, k_{56} ,
\end{eqnarray}
and
\begin{eqnarray}
 (\epsilon_{45}, p-k_{45}) - \frac{1}{2}(\epsilon_4,\epsilon_5)   & = & - (\eta
 \tilde\lambda_4, p - k_{45}) + \frac{1}{2} \,
 \frac{\langle\eta\, 5\rangle}{\langle4\, 5\rangle} \, (I^{(4)} - I^{(6)}) , \\
(\epsilon_{56}, p-k_{4}) - \frac{1}{2}(\epsilon_5,\epsilon_6)   &
= &   (\eta
 \tilde\lambda_6, p - k_{4}) - \frac{1}{2} \,
 \frac{\langle\eta\, 5\rangle}{\langle6\, 5\rangle} \, (I^{(5)} - I^{(*)}) ,
\end{eqnarray}
where $I^{(4)} = p^2$, $I^{(5)} = (p-k_4)^2$, $I^{(6)} =
(p-k_{45})^2$ and $I^{(*)} = (p-k_{456})^2$  are inverse
propagators.

Now considering the three terms coming from the first 3 Feynman
diagrams shown in Fig.~\ref{Five}, we have:
\begin{eqnarray}
A_{456} & = & (\epsilon_4, p-k_4)(\epsilon_5,p-k_4)(\epsilon_6,p-k_4) \nonumber \\
&  + & ((\epsilon_{45}, p-k_{45}) -
\frac{1}{2}(\epsilon_4,\epsilon_5)) (\epsilon_6, p-k_4)
 \, I^{(5)} \nonumber \\
&  + & (\epsilon_4, p-k_4) ((\epsilon_{56}, p-k_{4}) -
\frac{1}{2}(\epsilon_5,\epsilon_6)) \, I^{(6)}
\nonumber \\
&  = & ((\lambda_5\tilde\lambda_4, p-k_4)(\eta\tilde\lambda_6,
p-k_4) -
 (\eta\tilde\lambda_4, p-k_4) (\lambda_5\tilde\lambda_6,p-k_4)) \, I^{(6)}
 \nonumber \\
& + & \frac{1}{2} \frac{\langle\eta\, 5\rangle}{\langle4\,
5\rangle} \, \, (\lambda_5\tilde\lambda_6,p-k_4)\, (I^{(4)} -
I^{(6)}) \, I^{(5)}
\nonumber \\
& - & \frac{1}{2} \frac{\langle\eta\, 5\rangle}{\langle6\,
5\rangle} \, \, (\lambda_5\tilde\lambda_4,p-k_4)\, (I^{(5)} -
I^{(*)}) \, I^{(6)} ,
\end{eqnarray}
by doing tensor reduction with $k_{45}$ for the first term. This
can be further simplified by expressing  $\eta$ in terms of a
linear combination of $\lambda_{4,5}$. The final result is:
\begin{eqnarray}
A_{456} & = & \frac{1}{2} \frac{\langle\eta\, 5\rangle}{\langle4\,
5\rangle} \, \, (\lambda_5\tilde\lambda_6,p-k_4)\,  I^{(4)}   \,
I^{(5)} +  \frac{1}{2} \frac{\langle\eta\, 5\rangle}{\langle6\,
5\rangle}
 \, (\lambda_5\tilde\lambda_4,p-k_4)\,   I^{(*)}  \, I^{(6)}
 \nonumber \\
 & +  & I^{(5)} \,  I^{(6)}  \, \langle\eta\, 5\rangle   \,
 \left[ [4\,6] - \frac{1}{2}   (\lambda_5( {\tilde\lambda_4\over \langle6\, 5\rangle} +
 {\tilde\lambda_6\over \langle4\, 5\rangle}) , p-k_4) \right] \nonumber \\
 & = &
 \frac{1}{2} \frac{\langle\eta\, 5\rangle}{\langle4\, 5\rangle}
\,   (\lambda_5\tilde\lambda_6,p-k_4)\,  I^{(4)}   \, I^{(5)} +
\frac{1}{2} \frac{\langle\eta\, 5\rangle}{\langle6\, 5\rangle}
 \, (\lambda_5\tilde\lambda_4,p-k_4)\,   I^{(*)}  \, I^{(6)}
 \nonumber \\
 & +  & \frac{1}{2} \, I^{(5)} \,  I^{(6)}  \, \langle\eta\, 5\rangle   \,
 \left[ [4\,6] +  \frac{1}{\langle4\, 5\rangle  \langle6\, 5\rangle}
   (\lambda_5 \tilde\lambda_{k_{456}}^{(5)}, p ) \right] ,
   \label{adjacentthree}
\end{eqnarray}
which has a nice symmetric property under the flipping operation $4
\leftrightarrow 6$. The last term in eq.~(\ref{adjacentthree})
actually cancels the contribution from the last Feynman diagram in
Fig.~\ref{Five}.

For the case of different neighboring helicities, we can use the
following reduction formulas:
\begin{eqnarray}
(\lambda_1\tilde\eta, p) \, (\lambda_1\tilde\lambda_2, p) & = &
{\langle\lambda_1 |p\,k_2\, K\, k_1\, p|\tilde\eta\rangle \over
\langle 2 |K | 1 \rangle } \nonumber \\
& & \hskip -2cm = {1\over \langle 2 |K_4| 1 \rangle } \, \Big(
I^{(1)} \, ( \lambda_1\tilde\lambda_2, p)\,
 \langle2|(K_4+k_1)| \tilde\eta\rangle   \nonumber \\
&  &  \hskip -2cm + I^{(2)} \, \langle 1|(p-k_2)(K_4k_1 -
k_2K_4)|\tilde\eta\rangle - I^{(3)} \, \langle 1| p  \,
K_4|1\rangle [1\, \tilde\eta]   \nonumber \\
& & \hskip -2cm - I^{(4)} \, (\lambda_1\tilde\lambda_2,p)\,
\langle2\, 1\rangle [1\, \tilde\eta] + (K_4+k_1)^2\,
(\lambda_1\tilde\lambda_2, p)\,  \langle2\, 1\rangle [1\,
\tilde\eta]\Big),    \label{boxtensora}
\\ (\lambda_1\tilde\lambda_2, p) \, (\eta\tilde\lambda_2, p) & = &
{\langle\eta |p\,k_2\, K\, k_1\, p|\tilde\lambda_2\rangle \over
\langle 2 |K | 1 \rangle } \nonumber \\
& & \hskip -2cm =  {1\over \langle 2
 |K_3| 1 \rangle } \, \Big(  I^{(1)} \, \langle\eta| k_2\,K_3\, p | 2
\rangle \nonumber \\
&  &  \hskip -2.5cm  + I^{(2)} \, \langle \eta|(K_3k_1
-k_2K_3)(p+k_1)|2\rangle  - I^{(3)}\, \langle \eta|(k_2 +
K_3)|1\rangle (\lambda_1\tilde\lambda_2, p) \nonumber \\
& & \hskip -2.5cm + I^{(*)} \, (\lambda_1\tilde\lambda_2,p)\,
\langle\eta\, 2\rangle [2\, 1] - (k_2 + K_3)^2\,
(\lambda_1\tilde\lambda_2, p)\,  \langle\eta\, 2\rangle [2\, 1]
\Big), \label{boxtensor}
\end{eqnarray}
where $I$'s are the various inverse propagators:
\begin{eqnarray}
& & I^{(1)} = (p+k_1)^2, \qquad I^{(2)} = p^2, \qquad I^{(3)} =
(p-k_2)^2, \\
& & I^{(4)} = (p+k_1+K_4)^2, \qquad I^{(*)} = (p-k_2-K_3).
\end{eqnarray}
In eqs.~(\ref{boxtensora}) and (\ref{boxtensor}), the momentum $K$
can be chosen as one of the nearby momenta to avoid the spurious
pole associated with a composite momentum. For two-mass-hard box, $K$
can only be chosen as one of the composite momenta.

Because of the complexity of the above tensor reduction formula
for different neighboring helicities, it is better not to use them directly.
Luckily we are able to avoid using them directly for tensor
reduction with 5- and 6-point diagrams by a judicious choice of
reference momenta for all the polarization vectors. The details
will be given in \cite{xyziii}. For two-mass-hard box integrals, we
must use the above general tensor reduction in order to obtain
comparatively compact analytic expressions. We will use a slightly
different reduction formula in Sect.~7.4 to compute the rational
part of the two-mass-hard box integral.

The above is just the first step for the tensor reduction. Of
course this procedure can be applied recursively. By a rough
examination of this recursive method, one immediately finds two
problems: 1) the association of the resulting polarization vector
with an external (composite) momentum may not satisfy the physical
conditions; 2) the above formula is no-longer applicable if one
has a massive external momentum (a composite one arising from the
reduction or a pinched line by sewing the tree to the loop). There are
other methods to do further tensor reduction, some involving Gram
determinants. Exactly because of these problems, tensor reduction
is usually the most difficult part and the bottleneck for directly
computing the one loop amplitude. Quite elaborate methods are
developed to tackle these problems. See, for example,
\cite{Denner,BinothZ}.

Because of the complexity of doing further tensor reduction, we
immediately see the problem why directly computing the amplitude
by using Feynman rules is an extraordinarily difficult task for
higher point amplitudes. There are mainly two difficulties to
overcome:  too many diagrams and the complexity for tensor
reduction (especially at a later stage of tensor reduction as
mentioned in the above). By using computers it is not too difficult
to manage the numerous Feynman diagrams (of the order 1000). But the
complexity of tensor reduction with the appearance of spurious
poles is actually the bottleneck for analytic computation. Tensor
reduction is also the bottleneck for doing numerical calculations.
See, for example, \cite{GRACEFeynman}.

In contrast we also see that why it is possible to compute the
rational part by using the conventional Feynman integrals. First
by using the supersymmetric decomposition, the number of Feynman
diagrams is about  50 (1 hexagon, 6 pentagons, 15 boxes, 20
triangles and 15 bubbles) by only computing the scalar loop
contributions. Second by computing only the rational part, it is
not necessary to do tensor reduction all the way down to scalar
integrals. One needs only to do tensor reduction to reduce the
degree by 2 (see next section). So tensor reduction does not
complicate the analytic expressions  significantly. In fact for
special helicity assignments, there is an almost mutual
cancellation between higher point diagrams and lower point
diagrams, as we demonstrated in eqs.~(\ref{eqreduction}) and
(\ref{eqreductiona}). This is a manifestation of gauge invariance.
This property can be used to check and to organize the results of
our calculation. It is very important to have some ``local''
cancellations before adding all the results together in order to
obtain relatively compact analytic results for the rational parts
of QCD amplitudes.

\section{The BDDK theorem and the structure of one-loop amplitudes}

\begin{figure}[ht]
\centerline{\includegraphics[height=4cm]{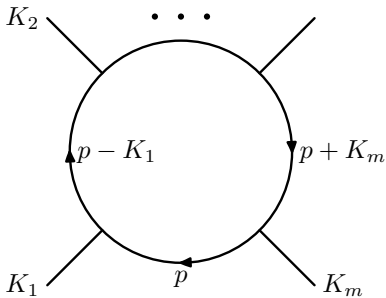} } \caption{A
generic one-loop diagram with external momenta $K_1$, $\cdots$,
$K_m$. $p$ is the internal momentum between the external lines
$K_1$ and $K_m$.} \label{Six}
\end{figure}

A generic $m$-point one-loop Feynman diagram shown in
Fig.~\ref{Six} and its integral is given as follows (by using
Feynman rules):
\begin{equation}
I_m^D[f(p)] =  \int { {\rm d}^D p \over i\,\pi^{D/2}} \, { f(p)
\over p^2 (p-K_1)^2 \cdots (p+K_m)^2} ,
\end{equation}
where $f(p)$ is a polynomial function of the internal momentum
$p$. For phenomenologically interesting models and by choosing a
suitable gauge, the degree of $f(p)$ is always not greater than
$m$. $f(p)$ also depends on the external momenta $k$ ($K_i$'s are
sums of cyclicly consecutive external momenta $k$'s) and the
polarization vectors $\epsilon_i$. For $f(p)=1$ it is called the
scalar integral. The strategy of computing $I_m^D[f(p)]$ is to
reduce it recursively into lower degree polynomials and/or lower
point integrals. It is a well-known result that $I_n^D[f]$ can be
generically written as:
\begin{eqnarray}
I_m^D[f] & = & \sum_i c_{4,i}(\epsilon,k; D) \, I_4^{D(i)}[1]
\nonumber
\\
& + & \sum_i c_{3,i}(\epsilon,k; D) \, I_3^{D(i)}[1] + \sum_i
c_{2,i}(\epsilon,k; D) \, I_2^{D(i)}[1], \label{imfp}
\end{eqnarray}
up to infinitesimal terms arising from tensor reduction from
pentagon or higher point diagrams. Here $c_{j,i}$'s are rational
functions of the external momenta when all polarization vectors
are written in terms of spinor products. We note that these
coefficients also depend on the (arbitrary) space-time dimension
$D$ in dimensional regularization (in the FDH scheme \cite{FDH}).

A brute-force computation of these coefficients from Feynman
integrals is  an impossible task for 6 or higher point amplitudes.
The 5-point case was computed by string-inspired method by using a
table for all Feynman parameter integrals (see below)
\cite{BernKosower}. However the string-inspired method is still
not powerful enough to compute even the 6-gluon amplitude due to
the complexity of the Feynman integrals and the intermediate
expressions.

For physical application what we actually need is an expansion in
$\epsilon$  of the above formula  for $D= 4 - 2 \epsilon$, up to
finite terms. If we forget the infrared divergence for the moment,
there are only simple pole  ($1\over\epsilon$) terms in the scalar
integrals $I_{4,3,2}^D[1]$ with rational coefficients. So we can
write the $n$-gluon amplitude (addition of all contributing
$I_m^D[f(p)]$ from all Feynman diagrams) as follows:
\begin{eqnarray}
{\cal A}_n & = & \sum_i c_{4,i}(\epsilon,k; 4) \, I_4^{D(i)}[1]
+  \sum_i c_{3,i}(\epsilon,k; 4) \, I_3^{D(i)}[1] \nonumber \\
& + & \sum_i c_{2,i}(\epsilon,k; 4) \, I_2^{D(i)}[1] +
(\hbox{rational function}) + O(\epsilon) . \label{namplitude}
\end{eqnarray}

For supersymmetric  theories,  Bern, Dixon, Dunbar and Kosower
\cite{BDDK} proved  a theorem which states that the rational
function is exactly zero.  This is due to the better ultra-violet
behaviour of one-loop amplitudes in supersymmetric theories.
So we need only to compute the rational coefficients (called the
cut-constructible part hereafter) exactly at $D=4$.  What
they proved is actually a more general theorem: if $f(p)$ is a
polynomial (in $p$) of degree $m-2$ or less, the rational part for
$I_m^D[f(p)]$ arising by expanding in $\epsilon$ is exactly zero.
Generally speaking, the rational part is non-vanishing for degree
$m$ and $m-1$ polynomials.

For a non-supersymmetric theory like QCD we also need to compute the
rational function (called the rational part hereafter). In a
series of papers \cite{BDDK,BDKB,BDK}, Bern, Dunbar, Dixon and
Kosower have developed a method of computing  the
cut-constructible part, i.e.,  the coefficients $
c_{j,i}(\epsilon,k; 4)$,  from 4-dimensional unitarity. The nicety
of 4-dimensional unitarity is that all the ingredients in the
unitarity relation are on-shell quantities. However 4-dimensional
unitarity loses all information about the rational function part
in eq.~(\ref{namplitude}). One must use some other methods to
compute the rational part. In \cite{BDKB}, they use the
factorization properties with  trial and error. The difficulty
of computing the rational function prevents the wider application
of the unitarity method to calculate more general amplitudes. As we
mentioned in the introduction, the rational function part could be
computed by going to $D$-dimensional unitarity \cite{BDK, AMST}.
%but  too much information is needed and it is even harder.

In our papers, the precise definition of the rational part $R_n$ of the QCD amplitude
 is as follows.
The $n$-point one loop  color-ordered partial amplitude $A_{n;1}^{[0]}$ defined in eq. (2.10)
in the second paper of \cite{BDKBoot} can be decomposed as
\begin{eqnarray}
A_{n;1}^{[0]} & = &{(4\pi)^\epsilon\over 16\pi^2}\bigg( \sum_i c_{4,i}(\epsilon,k; 4) \, I_4^{D(i)}[1]
+  \sum_i c_{3,i}(\epsilon,k; 4) \, I_3^{D(i)}[1] \nonumber \\
& + & \sum_i c_{2,i}(\epsilon,k; 4) \, I_2^{D(i)}[1] +
2 R_n + O(\epsilon) \bigg)\,,
%\\ r_\Gamma & = & {\Gamma(1+\epsilon)\Gamma^2(1-\epsilon)\over \Gamma(1-2\epsilon)}\,,
\end{eqnarray}
where $R_n$ is the rational part for one real scalar circulating in the loop.
% Since $r_\Gamma=1+O(\epsilon)$, it can also be omitted.
We will use notations like $R(--++++)$ to denote $R_n$ in different cases in
 \cite{xyzii} and \cite{xyziii}.

In the following sections we will exploit the BDDK theorem to
compute the rational part directly from the Feynman integrals.
By using the recursive relations satisfied by the tensor integrals we
will derive the recursive relations for the rational parts by
making an expansion in $\epsilon$.  Our integration method of
computing the rational part may also be used to compute the
rational part by using $D$-dimensional unitarity.
 We use  then
recursive relations  to derive explicit formulas for the rational
parts of all bubble and triangle integrals in Sect.~6. In Sect.~7,
we derive the formulas for box integrals up to two-mass-hard boxes.
Formulas for 3-mass and 4-mass box integrals can also be derived.
%and will be given elsewhere.
They are much more complicated than
the two-mass-box formulas. Fortunately they are not needed in the
computation of 6-gluon amplitudes and will not be given here.
We note that the recursive
relations for tensor integrals and the rational parts can also be
derived for massive internal loop and/or external fermion lines.
For simplicity all formulas are given only for the cases with vanishing internal
masses.

\section{The recursive relations of one-loop amplitudes}

In this section we study the recursive relations of one-loop
amplitudes \cite{BDKReduction,BinothZ,Denner}. By using Feynman
parametrization we have
\begin{eqnarray}
I_n^D[1] & \equiv &  \int { {\rm d}^D p \over i  \pi^{D/2}} \, {
1 \over p^2 (p-k_1)^2 \cdots (p+k_n)^2}  \nonumber \\
& = & (-1)^n\, \Gamma(n-D/2) \, \int { {\rm d}^n a} \, { \delta( 1
- \sum_i a_i) \over ( a \cdot S \cdot a )^{n -{D\over 2}} } ,
\end{eqnarray}
where
$$a \cdot S \cdot a  = \sum_{i,j=1}^n a_i\,a_j\, S_{ij}$$
and the matrix $S$,
\begin{equation}
S =  - {1\over 2} \, \left(
\begin{array}{ccccc}
0 & k_1^2 & (k_1+k_2)^2 & \cdots & (k_1+k_2+\cdots k_{n-1})^2 \cr
* & 0 & k_2^2 & \cdots & (k_2+k_3+\cdots k_{n-1})^2 \cr
\vdots & \vdots & \vdots & \vdots & \vdots \cr
* & * & * & 0  &     k_{n-1}^2 \cr
* & * & * & *  & 0
\end{array}
\right) ,
\end{equation}
is an $n\times n$ symmetric matrix of external kinematic variables
(extension to massive loop is straightforward). For tensor
integral $I_n^D[f(p)]$ it is given by the Feynman parameter integral with an
extra polynomial of $a$ in the numerator\footnote{We note that
the difference between Feynman parameter integral $\hat{I}_n[g(a)]$
and $I_n[f(p)]$ is only the $(-1)^n$ factor. For even $n$ and $f(p)=g(a)=1$,
they are the same and we will not distinguish them in this case.
}:
\begin{eqnarray}
\hat{I}_n^D[g(a)] =  \Gamma(n-D/2) \, \int { {\rm d}^n a} \, { \delta( 1
- \sum_i a_i) \, g(a)  \over ( a \cdot S \cdot a )^{n -{D\over 2}} } .
\end{eqnarray}
The degree of $g(a)$ is the same as the degree of $f(p)$ in $p$.

As explained in \cite{BDKReduction},
a set of recursive relations for these tensor integrals  can be
 derived by performing the following integration:
\begin{eqnarray}
F & = & \Gamma(\alpha) \int_0^1 {\rm d} a_{n-1} \int_{0}^{1-a_{n-1}} {\rm d}
a_{n-2} \cdots \int_0^{ 1- a_1-a_2 - \cdots- \hat a_m - \cdots - a_{n-1}} {\rm d} a_m \,
\cdots \nonumber \\
&  & \times \, {\partial \over \partial a_m}\left[ \left.
{f(a) \over ( a\cdot S \cdot a )^\alpha}\right|_{a_n = 1- a_1 -\cdots - a_{n-1}}
\right] ,
\end{eqnarray}
in two ways: one by partial integration and one by direct differentiation. Then we
obtain the following recursive relation:
\begin{eqnarray}
& & \hskip-2cm
- 2 \Gamma(\alpha+1) \,\int_0^1{\rm d}^na \delta(1-a) \, { f(a)( (S\cdot a)_m -
(S\cdot a)_n ) \over (a\cdot S\cdot a)^{\alpha+1} }  \nonumber \\
& = & \Gamma(\alpha) \int_0^1 {\rm d}^na \delta(1-a)\, \left. {f(a) \over
 (a\cdot S\cdot a)^{\alpha } }\right|_{a_n=0} \nonumber \\
 & & -
 \Gamma(\alpha) \int_0^1 {\rm d}^na \delta(1-a)\, \left. {f(a) \over
 (a\cdot S\cdot a)^{\alpha } }\right|_{a_m=0} \nonumber \\
 & & - \Gamma(\alpha)\,
 \int_0^1 {\rm d}^na \delta(1-a)\,   {\partial_m f(a) - \partial_n f(a) \over
 (a\cdot S\cdot a)^{\alpha } }  .
\end{eqnarray}
By carefully examining the composition of $  a\cdot S\cdot a $ for
$a_m=0$ one recognizes that this corresponds to a pinching limit
of the original Feynman diagram: $k_{m-1},~k_m \rightarrow
k_{m-1}+k_{m}$. By setting $\alpha = n-1-{D\over2}$ and using the
definition for $\hat{I}_n^D[f]$, the above recursive relation translates
into the following form:
\begin{eqnarray}
& & \hskip -1cm \hat{I}_n^D[-2f(a)( (S\cdot a)_m - (S\cdot a)_n)] \nonumber \\
 &  =  &
\hat{I}_{n-1}^{D(n)}[f(a)] -  \hat{I}_{n-1}^{D(m)}[f(a)] -
\hat{I}_n^{D+2}[\partial_mf(a)]  + \hat{I}_n^{D+2}[\partial_nf(a)]  .
\label{recursiveBernA}
\end{eqnarray}
Here $\hat{I}_{n-1}^{D(i)}[f(a)]$ denotes the $(n-1)$-point one-loop Feynman integral obtained
by pinching $k_{i-1}$ and $k_{i}$.
Coupled with one more equation from the delta function:
\begin{equation}
\sum_{i=1}^n \hat{I}_n^D[ f(a) a_i ] = \hat{I}_N^D[f(a)],
\label{recursiveBernB}
\end{equation}
we can solve $\hat{I}_n^D[ f(a) a_i]$ in terms of $\hat{I}_{n-1}^{D(m)}[f(a)]$,
$\hat{I}_{n}^{D+2}[\partial_mf(a)]$ and $\hat{I}_{n}^{D }[f(a)]$:
\begin{eqnarray}
\hat{I}_n^D[ f(a) a_i ] & = & c_{ij}^{(n)} \hat{I}_{n-1}^{D(j)}[ f(a) ] +
d_{ij}^{(n)} \hat{I}_n^{D+2}[ \partial_j f(a)] + c_i^{(n)}\hat{I}_n^D[f(a)] .
\end{eqnarray}
The above reasoning goes through so long as we can invert the relevant matrix. This matrix
is
\begin{equation}
\left[
\begin{array}{ccc}
& -2(S_{ij} - S_{nj}) & \cr
1 & \cdots & 1
\end{array}
\right]
\end{equation}
and it is singular for $n\ge 6$. For our purpose of computing the
rational part we will only need these recursive relations for
$n\le4$. Higher point tensor integrals are reduced directly in
$D=4$ as we have done in the last section. For further discussions
about the tensor reduction and its close relation with the above
recursive relation, we refer the reader to \cite{BinothZ,Denner}.

The above recursive relation is not symmetric for all $a_i$. The symmetric recursive
relation can be obtained by firstly solving $\hat{I}_n^D[f(a)(S\cdot a)_n]$ and substituting
it back to the system of equations. By taking $f= g_l(a)\, a_m$
($g_l(a)$ is a homogeneous polynomial of degree $l$ in $a$), we can
solve $\hat{I}_n^D[g_l(a)\, (S\cdot a)_n]$ to get:
\begin{equation}
2 \hat{I}_n^D[ g_l(a)\,(S\cdot a)_n] = \hat{I}_{n-1}^{D(n)}[g_l(a)] +
(n-1-l-D)\, \hat{I}_n^{D+2}[g_l(a)] + \hat{I}_n^{D+2}[\partial_ng_l(a)] .
\end{equation}
Substituting this back into eq.~(\ref{recursiveBernA}) and
multiplying both sides of the equation by $S^{-1}_{i\,m}$ and then
summing over $m$, we have \cite{BinothZ}:
\begin{eqnarray}
\hat{I}_n^D[g_l(a)\, a_i] & = & {1\over2}\, (n-1-l-D)\,\gamma_i\, \hat{I}_{n}^{D+2}
[g_l(a)] \nonumber \\
& + &
{1\over2}\, \sum_j {S_{ij}^{-1}}\, \hat{I}_{n-1}^{D(j)}[g_l(a)] +
{1\over2}\, \sum_j {S_{ij}^{-1}}\, \hat{I}_{n}^{D+2}[\partial_j g_l(a)],
\label{higherD}
\end{eqnarray}
where
\begin{equation}
\gamma_i =  \sum_j S_{ij}^{-1} .
\end{equation}
One can check that the above recursive relation  is equivalent to the
following recursive relation:
\begin{eqnarray}
\hat{I}_n^{D }[ a_i f(a) ] & = & P_{ij}\, \left( \hat{I}_n^{D (j)}[f(a)] +
\hat{I}_n^{ D+2 } [\partial_jf(a) ] \right) +
{\gamma_i\over \Delta} \, \hat{I}_n^D[f(a)] ,  \label{recursivesymm}\\
P_{ij} & = &
{1\over 2}\, \left( S_{ij}^{-1} - {\gamma_i\, \gamma_j \over \Delta }\right),
\qquad
\Delta = \sum_i\gamma_i .
\end{eqnarray}
The good point of this recursive relation is that all coefficients
have no explicit dependence on the space-time dimension $D$ and so
it is well suited to compute the rational part.

We note that the above reasoning would go through if $S$ is invertible.
In the two-mass and one-mass
triangle cases,  $S$ is not invertible.  Nevertheless we can still
use the above formulas by taking a limit from the general 3-mass
triangle case. The point is that all we need is to have a
well-behaved limit for the quantity $P_{ij}$ and ${\gamma_i\over
\Delta}$. We have verified that directly taking the massless limit
gives the same result as one would obtained by solving the
non-singular system of equations (\ref{recursiveBernA}) and
(\ref{recursiveBernB}).

From eq.~(\ref{recursivesymm}) we see that higher dimensional
integrals also appear in the recursive relations. These higher
dimensional tensor integrals can  be reduced to even higher and/or
lower point tensor integrals by using these recursive relations
repeatedly. At the end only scalar integrals are left. These
higher dimensional scalar integrals can be reduced to lower
dimensional and lower point scalar integrals by using the
following  recursive relation:
\begin{equation}
\hat{I}^{D+2}_n[1] = {1\over (n-1-D)\, \Delta} \, \left[ 2 \, \hat{I}_n^D[1] -
\sum_j
 \gamma_j \, \hat{I}_{n-1}^{D(j)}[1] \right]  . \label{higherDD}
\end{equation}
This equation is derived from eq.~(\ref{higherD}) by setting
$g_l(a)=1$ and summing over $i$. The explicit dependence on the
space-time dimension $D$ in eq.~(\ref{higherDD}) is very
important. Otherwise all the rational coefficients in
eq.~(\ref{imfp}) would have no  explicit dependence on $D$ and there
would be no rational part.

To be more specific, we give some explicit examples for bubble and
triangle integrals in  what follows.

For bubble integral we have
\begin{equation}
I_2^D[ f(p) ] \equiv  \int { {\rm d}^D p \over i \, \pi^{D/2} } \,
{ f(p) \over p^2\, (p + K)^2} ,
\end{equation}
where $K$ is the sum of momenta on one side of the bubble diagram.
For $K^2=0$ this integral is 0 in dimensional regularization. So
we will assume $K^2\neq 0$ hereafter. By direct computation we
have the following results ($D=4-2\epsilon$):
\begin{eqnarray}
I_2^D[ 1 ] & = & {r_\Gamma \over \epsilon(1-2\epsilon) }\, (-K^2)^{-\epsilon},\hspace{1cm}
r_\Gamma={\Gamma(1+\epsilon)\Gamma^2(1-\epsilon)\over \Gamma(1-2\epsilon)}\,,
\\
I_2^D[p^\mu] & = & - {K^\mu \over 2}\, I_2^D[ 1 ] , \\
\hat{I}_2^D[ a_1^2 ] & = & \hat{I}_2^D[a_2^2] = {2  - \epsilon \over
2(3-2\epsilon) } \, I_2^D[1] . \label{bubbleaa}
\end{eqnarray}

The recursive relation (\ref{higherDD}) becomes:
\begin{equation}
I_2^{D+2}[1]   =   {K^2 \over 2(D-1)}\, I_2^D[1]  .
\label{bubbleD}
\end{equation}
This  can be applied recursively to compute arbitrarily  higher
dimensional bubble integrals.

\begin{figure}[ht]
\centerline{\includegraphics[height=4cm]{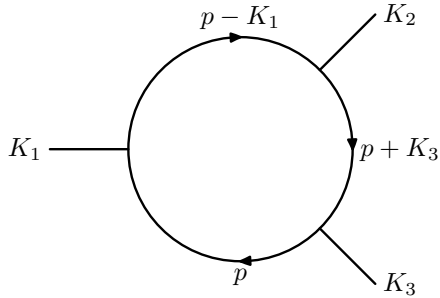} }
\caption{A generic three-mass triangle diagram. The 2-mass
triangle diagram is obtained by setting one of momentum to be
massless, for example $K_1=k_1$ and  $k^2_i=0$. } \label{Seven}
\end{figure}

A generic triangle diagram is shown in Fig.~\ref{Seven} and the
integral is\footnote{A  minus sign is not included in the
definition of $I_3^D[ f(p) ]$.}
\begin{equation}
I_3^D[ f(p) ] \equiv   \int { {\rm d}^D p \over i \, \pi^{D/2} }
\, { f(p) \over p^2\, (p - K_1)^2\, (p+K_3)^2 } . \label{ithreemass}
\end{equation}
This integral is finite (free of ultra-violet (for degree 3 or
less polynomial $f(p)$) and infrared divergences) for generic
external momenta (i.e., $K_i^2\neq 0$, $i=1,2,3$). The explicit
formula can be found in \cite{BDDK}. For degenerate cases we have
\begin{eqnarray}
I_3^D[1]  & = & - {1\over \epsilon^2} \,
{\Gamma(1-\epsilon)^2\Gamma(1+\epsilon)\over \Gamma(1-2\epsilon)}
\, { (-K_2^2)^{-\epsilon} - (-K_3^2)^{-\epsilon} \over (-K_2^2)  -
(-K_3^2) }, \quad K_1^2 = 0 ,
\\
I_3^D[1]  & =& -  {1\over \epsilon^2} \,
{\Gamma(1-\epsilon)^2\Gamma(1+\epsilon)\over \Gamma(1-2\epsilon)}
\, { (-K_3^2)^{-1-\epsilon}   }, \qquad K_1^2 = K_2^2 = 0 .
\end{eqnarray}

In the next section we will use these results and the more general
recursive relations to derive the rational parts of the bubble,
triangle and box integrals by making an expansion in the
space-time parameter $\epsilon$.

\section{The rational parts of the triangle integrals}

\subsection{$\epsilon$-expansion and the rational parts:
the bubble integrals}

The $\epsilon$-expansion of the scalar bubble integral is:
\begin{eqnarray}
I_2^D[1] & = & {1\over \epsilon} + O(1) .
\end{eqnarray}
By using this result in eq.~(\ref{bubbleD}) we have
($D=4-2\epsilon$):
\begin{eqnarray}
I_2^{D+2}[1] & = & {K^2 \over 6}\, I_2^D[1] + \, {K^2\over 9}  + O(\epsilon), \\
I_2^{D+4}[1] & = & {(K^2)^2 \over 60}\, I_2^D[1] + \, {4
\,(K^2)^2\over 225} + O(\epsilon) .
\end{eqnarray}
We note that the first term on the right-hand of the above two
equations still depends on $D$  through the scalar integral $
I_2^D[1]$. What is important is the second term which is a pure
rational function and does not depend on the space-time dimension
$D$. This is the rational function part we need to keep track of
later.

By making an expansion in $\epsilon$ for eq.~(\ref{bubbleaa}), we
have:
\begin{equation}
\hat{I}^D_2[a_1^2] = \hat{I}^D_2[a_2^2] = {1\over 3} \, {I}^D_2[1] +   {1\over
18} + O(\epsilon).
\end{equation}
From this equation we read off the rational part of $
\hat{I}^D_2[a_1^2]$ and $\hat{I}^D_2[a_2^2]$ as $1\over18$. Translating back
to the momentum integral the result is:
\begin{eqnarray}
I_2^D[ (\epsilon_1,   p)\,(\epsilon_2 , p) ]  & = &  \left(
{(\epsilon_1,K)\,(\epsilon_2,K)\over 3} -
{K^2 \, (\epsilon_1,\epsilon_2) \over 6} \right)\, I_2^D[1]  \nonumber \\
& + & {1 \over 18} \, ( (\epsilon_1, K) \, (\epsilon_2, K) - 2 \,
K^2\, (\epsilon_1, \epsilon_2) ).
\end{eqnarray}
This result has already appeared in \cite{BDK} (eq.~(31) on
p.~133). We interpret the second term in the above equation as the
rational part. If we discard the first term we can simply write:
\begin{equation}
I_2[ (\epsilon_1,   p)\,(\epsilon_2, p) ] = {1 \over 18} \, (
(\epsilon_1, K) \, (\epsilon_2, K) - 2 \, K^2\, (\epsilon_1,
\epsilon_2) ),
\end{equation}
by dropping also the explicit dependence of $I_2$ on $D$.
However we  still retain this dependence for higher dimensional
Feynman integrals  and simply drop all the cut-constructible parts.
Explicitly we have:
\begin{eqnarray}
I_2[1] & = & 0 ,\\
\hat{I}_2[a_1^2] & = &  \hat{I}_2[a_2^2] = - \hat{I}_2[a_1\, a_2] = {1\over18}, \\
I_2^{D+2}[1] & = & {K^2\over 9}.
\end{eqnarray}

\subsection{The rational parts of the higher-dimensional scalar
integrals}

For higher dimensional scalar integrals we can use the recursive
relation eq.~(\ref{higherDD}) to derive their rational parts.

For three-mass triangle the explicit recursive relation is:
\begin{eqnarray}
\hat{I}_3^{3m\, (D+2)} [1]& = & {1\over (2-D) \, \Delta} \, \left[
2\,s_1\,s_2\,s_3\,\hat{I}_3^{3m } + s_2(s_1+s_3-s_2)\, I_2^{ (1)}  \right. \nonumber \\
&  + & s_3\,(s_1+s_2-s_3)\, I_2^{ (2)} +
\left.  s_1\,(s_2+s_3-s_1)\, I_2^{ (3)} \right]  , \\
\Delta & = &   s_1^2 + s_2^2 + s_3^2 -2(s_1\,s_2+
s_1\,s_3+s_2\,s_3) , \label{trianglerecursive}
\end{eqnarray}
where $s_i = K_i^2$, $i=1,2,3$.

For 3-mass triangle integral $I_3^D[1]$ is regular as $D\to 4$. By
using the $\epsilon$-expansion of $I_2^D$ we have:
\begin{eqnarray}
\hat{I}_3^{3m\,(D+2)}[1] & = & - {1\over 2 \, \Delta} \, \left[
2\,s_1\,s_2\,s_3\,\hat{I}_3^{3m } + s_2(s_1+s_3-s_2)\, I_2^{ (1)}  \right. \nonumber \\
& & \hskip -2cm +  s_3\,(s_1+s_2-s_3)\, I_2^{ (2)} + \left.
s_1\,(s_2+s_3-s_1)\, I_2^{ (3)} \right]  + {1\over2}   +
O(\epsilon). \label{trianglerational}
\end{eqnarray}
So the rational part of (the Feynman parameter integral)
$\hat{I}_3^{D+2}[1]$ is $ {1\over2}$. This result
also applies to the two-mass and one-mass triangle, although there are
intricacies of infrared divergences. This can be explicitly checked by
using the following explicit recursive relations for the
two-mass and one-mass triangle integrals:
\begin{eqnarray}
\hat{I}_3^{2m\, (D+2)}[1] & = & {1\over D-2}\, \left( {s_1\over s_1 - s_2}
\, I_2^{ (3)} -
{s_2\over s_1 - s_2} \, I_2^{ (1)} \right), \label{I32m}\\
\hat{I}_3^{1m\, (D+2)}[1] & = &  {1\over D-2}\, \, {I}_2^{ (3)} , \label{I31m}
\end{eqnarray}
and the explicit formulas for $I_2^{D(i)}$'s. We note that the
above recursive relations can be derived either by taking the
limit $s_1=K_1^2\to0$ (2-mass) or $s_{1,2}=K_{1,2}^2\to 0$ (1-mass)
 or by solving eqs.~(\ref{recursiveBernA}) and
(\ref{recursiveBernB}). In above eqs. (\ref{I32m}) and (\ref{I31m}), the terms with
infrared divergences really do not appear. This is because the
coefficients of the infrared divergent terms in eq. (\ref{trianglerecursive}) are
automatically zero. The infrared divergences also do
not contribute to the rational parts in the four point cases, which can be
checked by explicit calculations. A heuristic argument
is that since  higher dimensional integrals  are free
of infrared divergences, we expect that there
 should not be infrared divergence in the square bracket on the right hand side
of eq. (\ref{higherDD}) like in eq. (\ref{I32m}) and (\ref{I31m}), and hence the
infrared divergences do not contribute to the rational part.

The end result of the above analysis is
\begin{equation}
\hat{I}_3^{D+2}[1]  =  {1\over 2} + c(s) \, \hat{I}_3^D[1] +
\sum_i c_i(s) I_2^{D(i)}[1] + O(\epsilon),
\end{equation}
where $c(s)$ and $c_i(s)$'s are (rational) functions of the
external kinematic variables $s_i$ and the polarization vectors.
This formula  applies to all
possible triangle integrals. This shows explicitly that  the
rational part of the higher dimensional scalar integral
$\hat{I}_3^{D+2}[1]$  is a purely ultra-violet effect.

Setting the cut-constructible part to 0, we can effectively write
the following formula for the rational part:
\begin{equation}
\hat{I}_3^{D+2}[1]  = {1\over 2}.
\end{equation}

We have also  studied in detail the box integrals by using the
explicit formulas of \cite{BDDK}. We checked all the degenerate cases.
The results can be simply stated as follows:
\begin{eqnarray}
I_4^{D+2 }[1] & = & \hat{I}_4^{D+2}[a_i] = 0 , \\
I_4^{D+4 }[1] & = & {5 \over 18},
\end{eqnarray}
by dropping all the cut-constructible part. For 5-point integrals
we have
\begin{eqnarray}
\hat{I}_5^{D+2}[1]  &  = & \hat{I}_5^{D+2}[a_i] = \hat{I}_5^{D+2}[a_i\,a_j] =0, \\
\hat{I}_5^{D+4} [1] &  = & \hat{I}_5^{D+4} [a_i] = 0, \\
\hat{I}_5^{D+6}[1] & = & {13\over 144},
\end{eqnarray}
although they are not used in the computation of the QCD
amplitudes.

The rational part for the tensor integral is computed by first
transforming it into Feynman parameter integrals and then using the
recursive relations. The recursive relation for the rational part
is exactly the same as for the complete  Feynman integral. However
all lower degree ($m-2$ or less for $m$-point) Feynman integrals
can be set to zero by BDDK theorem. Effectively the recursive
relations are truncated.
%The general structure for the box and
%triangle recursive relations was shown in Fig.~5 and Fig.~6.
In the next few sections we compute explicitly the rational parts
for triangle and box integrals. From hereafter all Feynman
integrals will mean their rational parts, except explicitly stated otherwise.

\subsection{The triangle integrals: the general case}

For triangle integrals we will  consider 2 cases: the two-mass
triangle and the three-mass triangle, although the two-mass case can be
obtained simply by setting one of the masses to be 0. The reason for
doing this is that the formulas simplify greatly for the two-mass
triangle integral. For some computations one may only need the
simplified formulas (for example the 5-gluon amplitudes in
\cite{xyzii}). We will first give the formulas for the general
three-mass triangle integrals.

The three-mass triangle diagram is  shown in Fig.~\ref{Seven} and the
external momenta are denoted by $K_i$ ($i=1,2,3$). We  use the
convention of denoting a light-like momentum by a lower case $k$,
i.e. $k^2=0$.

We will give the rational parts for both the degree 3 and degree 2
polynomials\footnote{It is also possible to derive the rational
parts for higher degree polynomials but they have no practical
usage in application to computations in the electroweak theory and
QCD.}. First we make the following definitions:
\begin{eqnarray}
I_3(\epsilon_1, \epsilon_2, \epsilon_3 ) & \equiv &  \int { {\rm
d}^D p \over i \pi^{D/2} } \, {  (\epsilon_1 ,p)  \, (\epsilon_2 ,
 p-K_1)  \,
(\epsilon_3 ,p+K_3)  \over p^2 (p-K_1)^2 (p+K_3)^2 }  ,  \\
I_3 (\epsilon_1,\epsilon_2 ) & \equiv &  \int { {\rm d}^D p \over
i \pi^{D/2} } \, {   (\epsilon_1,p) \,  (\epsilon_2,p) \over p^2
(p-K_1)^2 (p+K_3)^2 } .
\end{eqnarray}
The actual computation of the rational part is done by using Feynman parametrization
and we have:
\begin{eqnarray}
I_3(\epsilon_1, \epsilon_2, \epsilon_3 ) & = &  - \hat{I}_3
[(\epsilon_1, a)\, (\epsilon_2,a)(\epsilon_3,a)] \nonumber \\
& + &  \sum_{i=1}^3 (\epsilon_i,\epsilon_{i+1})
\hat{I}_3^{D+2}[(\epsilon_{i+2},a) ] , \label{thrermassaaa}
  \\
I_3(\epsilon_1, \epsilon_2) & = &  - \hat{I}_3 [(\epsilon_1,a)\,
(\epsilon_2,a)  ] +  {1 \over 2}\,    (\epsilon_1,\epsilon_{2}),
\label{thrermassaaaa}
\end{eqnarray}
by using the previous result for $\hat{I}_3^{D+2}[1]$. In the above we
have used the following shortened notation:
\begin{eqnarray}
(\epsilon_1 , a) & = & (\epsilon_1 , K_1) \, a_2 - (\epsilon_1 ,
K_3) \, a_3,
\\
(\epsilon_2 , a) & = & (\epsilon_2 , K_2) \, a_3 - (\epsilon_2 ,
K_1) \, a_1,
\\
(\epsilon_3 , a) & = & (\epsilon_3 , K_3) \, a_1 - (\epsilon_3 ,
K_2) \, a_2,
\end{eqnarray}
where $a_i$'s are Feynman parameters as used in Sect.~5. In order
to give compact formulas for the various quantities appearing in
eqs.~(\ref{thrermassaaa}) and (\ref{thrermassaaaa}), we first
define the following functions:
\begin{eqnarray}
F_0(s_1,s_2,s_3) & \equiv & - \hat{I}_3 [ a_1 a_2 a_3] \nonumber \\
& = & {10 \, s_1 s_2 s_3 \over 3 \Delta^2 } + {
(s_1 + s_2  + s_3) \over 6 \Delta}, \\
F_1(s_1,s_2,s_3) & \equiv & - \hat{I}_3 [ a_1 a_2(a_2+ a_3)] \nonumber \\
& = & {5 \, (s_1 + s_2 - s_3)\, s_2 s_3  \over 3
\Delta^2 } + {(s_1 - s_3) \over 3 \Delta}, \\
F_2(s_1,s_2,s_3) & \equiv & - \hat{I}_3 [ a_2 a_3(a_3+a_1)] \nonumber \\
& = & {5 \, (s_2 + s_3 - s_1)\, s_3 s_1  \over 3
\Delta^2 } + {(s_2 - s_1) \over 3 \Delta}, \\
F_3(s_1,s_2,s_3) & \equiv & - \hat{I}_3 [ a_3 a_1 (a_1+a_2)] \nonumber \\
& = & {5 \, (s_3 + s_1 - s_2)\, s_1 s_2  \over 3 \Delta^2 } +
{(s_3 - s_2) \over 3 \Delta} .
\end{eqnarray}
where $s_i = K_i^2$ and $\Delta =s_1 ^2+s_2 ^2+s_3 ^2-2 (s_1 s_2 +
s_2  s_3+ s_3 s_1)$  is the ``Gram determinant" for the triangle
diagram.

By using these functions we have
\begin{eqnarray}
& & \hskip -1cm \hat{I}_3 [(\epsilon_1,a)\,(\epsilon_2, a)\,(\epsilon_3
, a) ] =
  F_0(s_1, s_2, s_3) ( (\epsilon_1, K_1) \, (\epsilon_2,
K_1) \, (\epsilon_3, K_2)   \nonumber \\
& &   + (\epsilon_1, K_3) \, (\epsilon_2, K_2)\, (\epsilon_3, K_2)
+(\epsilon_1, K_3) \, (\epsilon_2, K_1)
\, (\epsilon_3, K_3) \nonumber \\
& &    +(\epsilon_1, K_3) \, (\epsilon_2, K_1)\,(\epsilon_3, K_2
)-(\epsilon_1, K_1)\, (\epsilon_2, K_2)\, (
\epsilon_3, K_3)  ) \nonumber \\
& &    +  \sum_{i=1}^3 (\epsilon_1, K_i)\, (\epsilon_2, K_i)\,
(\epsilon_3, K_ i)  \,  F_i(s_1, s_2, s_3)
\nonumber \\
& &  + {1\over 2 \, \Delta} \Big( (s_1 - s_2 - s_3) \,
(\epsilon_1, K_1) \, (\epsilon_2, K_1) \, (\epsilon_3, K_3)
\nonumber \\
& & + \, (s_2 - s_3 - s_1) \, (\epsilon_1, K_1) \, (\epsilon_2,
K_2) \,  (\epsilon_3, K_2)
\nonumber \\
& & + \, (s_3 - s_1 - s_2) \, (\epsilon_1, K_3) \, (\epsilon_2,
K_2) \, (\epsilon_3, K_3)  \Big),
\end{eqnarray}
\begin{eqnarray}
\hat{I}_3^{D+2 }[ (\epsilon_1 , a)] & = & (\epsilon_1, K_1 - K_3) \left(
{7 \over 36} + { s_2(s_3 + s_1 - s_2) \over 12 \Delta}
\right) \nonumber \\
& & + (\epsilon_1, K_2)\, {(s_1-s_3) (s_3+s_1 - s_2) \over 12
\Delta} , \\
\hat{I}_3^{D+2 }[ (\epsilon_2 , a)] & = & (\epsilon_2, K_2 - K_1) \left(
{7 \over 36} + { s_3(s_1 + s_2 - s_3) \over 12 \Delta}
\right) \nonumber \\
& & + (\epsilon_2, K_3) \, {(s_2-s_1) (s_1+s_2 - s_3) \over 12
\Delta} , \\
\hat{I}_3^{D+2 }[ (\epsilon_3 , a)] & = & (\epsilon_3, K_3 - K_2) \left(
{7 \over 36} + { s_1(s_2 + s_3 - s_1) \over 12 \Delta}
\right) \nonumber \\
& & + (\epsilon_3, K_1) \, {(s_3-s_2) (s_2+s_3 - s_1) \over 12
\Delta} ,
\end{eqnarray}
and
\begin{eqnarray}
\hat{I}_3[(\epsilon_i , a) \, (\epsilon_j , a) ] & = &   {1\over 2\,
\Delta} \, \Big( s_1 \, ((\epsilon_i , K_2) \, (\epsilon_j , K_3)
+ (\epsilon_i , K_3) \, (\epsilon_j , K_2)) \nonumber \\
&   + &   s_2 \, ((\epsilon_i , K_3) \, (\epsilon_j , K_1)+
(\epsilon_i , K_1) \, (\epsilon_j , K_3)) \nonumber \\
&  & \hskip -1cm + s_3 \, ((\epsilon_i , K_1) \, (\epsilon_j ,
K_2) + (\epsilon_i , K_2) \, (\epsilon_j , K_1))  \Big)  .
\end{eqnarray}
We note that some formulas (and also the function $F_i$) in the above are related by
permutations. We purposely wrote down the complete formulas in order to see the pattern.

\subsection{The triangle integrals: the two-mass case}

For two-mass triangle we set $K_1 = k_1$.  Then we can derive the
simplified formulas in the two-mass triangle case by setting $s_1=0$
in the above formulas. Also we redefine $(\epsilon_3,a) $ to be
$(\epsilon_3,  k_1)\, a_2 - (\epsilon_3,K_3) \, a_3$. This
corresponds to the change of $(\epsilon_3, p+K_3)$ to
$(\epsilon_3, p)$. This gives a more symmetric form for the
result, as one can see from the Feynman integral representation by
doing the transformation $2 \leftrightarrow 3$. We list all the
explicit formulas here because they are used heavily in the
actual computation of the 6 particle amplitudes. This is necessary
to obtain compact analytic formulas for the rational part of the
amplitude. We have:
\begin{eqnarray}
\hat{I}_3 [(\epsilon_1, a)\, (\epsilon_2 , a)\,(\epsilon_3 , a) ] & = &
 {(s_2 + s_3) \over
6(s_2 - s_3)^2} \, (\epsilon_1 , K_2) \, (\epsilon_2 , k_1) \,
(\epsilon_3 , k_1)
\nonumber \\
& & \hskip -2cm + {(\epsilon_1 , K_2)\over 6(s_2 - s_3)} \, (
(\epsilon_2 , k_1)\, (\epsilon_3 , K_3) - (\epsilon_2 , K_2)\,
(\epsilon_3 , k_1)) .
\end{eqnarray}
\begin{eqnarray}
\hat{I}_3^{D+2}[(\epsilon_1, a)] & = & {1 \over 9} \, (\epsilon_1 , K_2), \\
\hat{I}_3^{D+2}[(\epsilon_2, a)] & = & -{7\over 36} \, (\epsilon_2 ,
k_1) + {1\over 9} \, (\epsilon_2 , K_2) - {(s_2+s_3)\, (\epsilon_2
, k_1)
\over 12(s_2 - s_3) }, \\
\hat{I}_3^{D+2}[(\epsilon_3 , a)] & = &  {7\over 36} \, (\epsilon_3 ,
k_1) - {1\over 9} \, (\epsilon_3 , K_3) - {(s_2+s_3)\, (\epsilon_3
, k_1) \over 12(s_2 - s_3) }.
\end{eqnarray}
\begin{eqnarray}
\hat{I}_3 [(\epsilon_i, a)\, (\epsilon_j , a) ] & = & - \, {(s_2 + s_3)
\over
2 (s_2 - s_3)^2} \, (\epsilon_i , k_1) \, (\epsilon_j , k_1) \nonumber \\
& - &  {((\epsilon_i , k_1) \, (\epsilon_j ,  K_2- K_3) +
(\epsilon_j , k_1) \, (\epsilon_i ,  K_2- K_3) )  \over 4 \, (s_2
- s_3) }    .
\end{eqnarray}
We note that in the last formula the double pole term is absent if
one of the polarization vectors is associated with the first
momentum $k_1$ and satisfies the physical condition $(\epsilon,
k_1)=0$.

By using the above results we have
\begin{eqnarray}
I_3 (\epsilon_1,\epsilon_2) & \equiv &   \int {{\rm d}^D p \over i
\pi^{D/2}} \,
  { (\epsilon_1, p) \, (\epsilon_2 , p) \,  \over p^2(p-k_1)^2
  (p+K_3)^2}    \nonumber \\
 & = &      {1\over 2}\, (\epsilon_1, \epsilon_2) +
   {(K^2_2 + K^2_3)\over 2 (K^2_2-K^2_3)^2 }
  \, (\epsilon_1, k_1) \, (\epsilon_2, k_1)  \nonumber \\
& + & {( (\epsilon_1, K_2)\,(\epsilon_2,k_1)  -(\epsilon_1, k_1)\,
(\epsilon_2, K_3) ) \over 2(K^2_2-K^2_3)}
 , \\
I_3 (\epsilon_1,\epsilon_2) & =  &
     {1\over2}\, (\epsilon_1, \epsilon_2) +  { (\epsilon_1, K_2) \,
(\epsilon_2, k_1)  \over 2(K_2^2-K^2_3)}  , \qquad (\epsilon_1,
k_1) = 0,
\end{eqnarray}
and
\begin{eqnarray}
I_3 (\epsilon_i) & \equiv &  \int {{\rm d}^D p \over i \pi^{D/2}}
\,\,
  { (\epsilon_1, p)  \, (\epsilon_2 ,  p-k_1) \, (\epsilon_3, p) \,
 \over p^2\, (p-k_1)^2 \,   (p+K_3)^2}
  \nonumber \\
  &   =  &   {1\over 36} \Big( (\epsilon_2,  4\,K_2 -7\, k_1)\,(\epsilon_1,
\epsilon_3) -(2 \leftrightarrow 3)  + 4 (\epsilon_1, K_2)\,
(\epsilon_2, \epsilon_3) \Big)  \nonumber \\
& - & {(K^2_2 + K^2_3)  \over 6\,(K^2_2-K^2_3)^2 } \, (\epsilon_1,
K_2) \,   (\epsilon_2, k_1) \, (\epsilon_3, k_1)
\nonumber \\
 & - & {(\epsilon_1, K_2)\, ((\epsilon_2, k_1)\,
(\epsilon_3, K_3) - (\epsilon_2 , K_2)\, (\epsilon_3, k_1))\over
6\, (K^2_2-K^2_3)}
\nonumber \\
& - & { (K^2_2 + K^2_3) \over 12\, (K^2_2-K^2_3) } \, (
(\epsilon_1, \epsilon_2)\, (\epsilon_3, k_1) + (\epsilon_1 ,
\epsilon_3)\, (\epsilon_2, k_1) )  .
\end{eqnarray}
The above formula is anti-symmetric under the exchange $2
\leftrightarrow 3$ by noting $(\epsilon_1, K_2) = -( \epsilon_1,
K_3)$. For later application we denote the two-mass rational part of
$I_3$ by $I_3^{2m(i)}$ where $i$ denote the massless external
line. To distinguish the two possible two-mass triangle diagrams
with the same massless external line  in the 6-gluon amplitude
case, we put a tilde on $I_3^{2m}$ for one of the two-mass triangle
integrals with 1, 3 and 2 external momenta in a clockwise
direction. Referring to Fig.~\ref{Seven}, $I_3^{2m(i)}$ has
external momenta $\{ k_i, k_{i+1}+k_{i+2},
k_{i+3}+k_{i+4}+k_{i+5}\}$, whereas $\tilde I_3^{2m(i)}$ has
external momenta $\{ k_i, k_{i+1}+k_{i+2}+k_{i+3},
k_{i+4}+k_{i+5}\}$.

\section{The rational parts of the box integrals}

\begin{figure}[ht]
\centerline{\includegraphics[height=4.5cm]{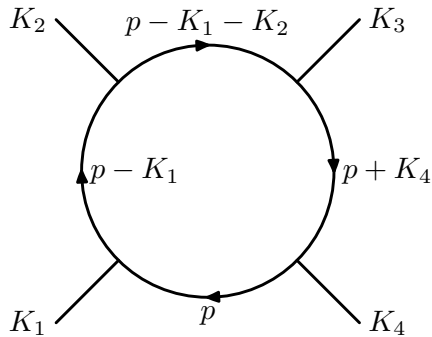} }
\caption{A generic box diagram. $p$ is the internal momenta
between $K_4$ and $K_1$. Other internal momenta are also shown
explicitly.} \label{Eight}
\end{figure}

A generic box diagram is shown in Fig.~\ref{Eight}. The kinematic
variables $s$ and $t$ are defined by following the standard
notation:
\begin{equation}
s = (K_1 + K_2)^2  = (K_3+K_4)^2, \qquad t =  (K_2 + K_3)^2  = (K_4+K_1)^2 .
\end{equation}
For our purpose of computing up to 6-gluon  amplitudes, we need to
consider up to two-mass boxes. There are two kinds of two-mass
boxes: the two-mass-hard box and two-mass-easy box. By a judicious
choice of reference momenta the two-mass-hard box does not show up in
the computation of MHV amplitudes.  So we will discuss only the two-mass
 case here and set $K_1 = k_1$, etc. by following the
convention of writing the light-like momenta in lower case $k$.
The one-mass box case is obtained as a special case of the two-mass-easy
 box case by setting further $K_2 = k_2$. We will list the
explicit formula for one-mass box for quick reference and the ease
of use.

\subsection{The box integrals of degree 3 polynomials: the two-mass-easy case}
Generally we need to compute the rational part of the following
box integral:
\begin{equation}
I_4 (\epsilon_1,\epsilon_2, \epsilon_3)   \equiv
 \int { {\rm d}^D \over i \pi^{D/2}}
\, { (\epsilon_1, p)  \, (\epsilon_2 , p) \, (\epsilon_3 , p)
\over p^2(p-K_1)^2 (p-K_1-K_2)^2 (p+K_4)^2}  .
\end{equation}

The complete  rational part in the general case is quite
complicated and should be avoided. From our experience it is
always the case that at least one of the polarization vectors
satisfies the physical condition for one of the massless external
momenta $k_1$ or $k_3$. In fact one can always expand an arbitrary
4-dimensional vector in terms of the 2 independent spinors of the
two external massless momenta. So we can assume that
$\epsilon_1$ satisfies the physical condition: $(\epsilon_1,
k_1)=0$.  To be specific we can take $\epsilon_1 = \eta\tilde{\lambda}_1$.
The negative helicity case can be obtained from this case (the
positive helicity one) by conjugation. If one of the polarizations
satisfies the physical condition for $k_3$ we can rotate the two-mass-easy
 box diagram by $\pi$ and relabel $k_3$ as $k_1$.

By explicit computation, we found that if the reference momentum
of $\epsilon_1$ is $k_3$, the rational part becomes quite simple
and is given as follows:
\begin{eqnarray}
I_4 (\lambda_3\tilde\lambda_1,\epsilon_2, \epsilon_3)  =
 {\langle 3 |K_2|1 ]   \over 2} \, \left[ { (\epsilon_2 , k_3) \,
(\epsilon_3 , k_3) \over (K_2^2 - t) (K_4^2 - s) } -{ (\epsilon_2
, k_1) \, (\epsilon_3 , k_1) \over (K_2^2 - s) (K_4^2 - t) }
\right] .
\end{eqnarray}
By using this result, the computation of the rational part of the
degree 3 polynomial can be proceeded by changing the reference
momentum of $\epsilon_1$ to $k_3$. This is equivalent to expanding
the spinor in terms of $\lambda_{1,3}$:
\begin{equation}
\eta = { \langle \eta \,  3\rangle \over \langle   1 \, 3 \rangle
}\, \, \lambda_1  +  { \langle \eta \,  1\rangle \over \langle 3\,
1 \rangle } \, \, \lambda_3.
\end{equation}
In so doing we also generate 2 triangle diagrams which have been
computed in last subsection. Explicitly we have:
\begin{eqnarray}
 I_4 (\eta \tilde\lambda_1,\epsilon_2,\epsilon_3 ) & = &  { \langle
\eta \,  1\rangle \over \langle 3\, 1 \rangle } \,
 {\langle 3 |K_2|1 ]   \over 2} \, \left[ { (\epsilon_2 , k_3) \,
(\epsilon_3 , k_3) \over (K_2^2 - t) (K_4^2 - s) } -{ (\epsilon_2
, k_1) \, (\epsilon_3 , k_1) \over (K_2^2 - s) (K_4^2 - t) }
\right]  \nonumber \\
& + & { \langle \eta \,  3 \rangle \over \langle   1 \, 3\rangle }
\, \left[ \tilde I_3^{2m} (\epsilon_2,\epsilon_3 ) - I_3^{2m}
(\epsilon_2,\epsilon_3 )\right] , \\
\tilde I_3^{2m}(\epsilon_2,\epsilon_3 ) & = &
 {1\over 2}\, (\epsilon_2, \epsilon_3) +
   {(K^2_2 + t)\over 2 (K^2_2- t)^2 }
  \, (\epsilon_2, k_3) \, (\epsilon_3, k_3)  \nonumber \\
& + & {( (\epsilon_2, k_3)\,(\epsilon_3,K_2)  -(\epsilon_3, k_3)\,
(\epsilon_2, K_4+k_1) ) \over 2(K^2_2-t)}, \nonumber \\
I_3^{2m} (\epsilon_2,\epsilon_3) & = &
 {1\over 2}\, (\epsilon_2, \epsilon_3) +
   {(K^2_4 + s)\over 2 (K^2_4-s)^2 }
  \, (\epsilon_2, k_3) \, (\epsilon_3, k_3)  \nonumber \\
& + & {( (\epsilon_2, k_3)\,(\epsilon_3,K_4)  -(\epsilon_3, k_3)\,
(\epsilon_2, k_1+K_2) ) \over 2(K^2_4-s)} .
\end{eqnarray}
 Taking
into account the anti-symmetric property of the product
$(\lambda_3\tilde\lambda_1, K_2)  = - (\lambda_3\tilde\lambda_1,
K_4)$, the above formulas are actually symmetric (which must be the
case) under the interchange $1\leftrightarrow 3$ and $2
\leftrightarrow 4$ (the polarization vectors are kept fixed
because they should be invariant under the permutation by themselves).

\subsection{The box integrals of degree 3 polynomials: the two-mass-hard case}

For the two-mass-hard box case, we follow the same strategy.  For
$\epsilon_1 = \lambda_1\tilde\lambda_2$ or
$\lambda_2\tilde\lambda_1$,  we have
\begin{eqnarray}
I_4^{2mh}(\lambda_1\tilde\lambda_2, \epsilon_2,\epsilon_3) & = & { \langle 1 | K_3 | 2 ]
\over 4 \, \delta } \,  I_4(\epsilon_2,\epsilon_3) , \\
I_4^{2mh}(\lambda_2\tilde\lambda_1, \epsilon_2,\epsilon_3) & = & { \langle 2 | K_3 | 1 ]
\over 4 \, \delta } \,  I_4(\epsilon_2,\epsilon_3) ,
\end{eqnarray}
where
\begin{eqnarray}
I_4(\epsilon_2,\epsilon_3)  & = & (\epsilon_2, k_1)(\epsilon_3,
K_4) + (\epsilon_2, K_4)(\epsilon_3, k_1) +
 (\epsilon_2, k_2) (\epsilon_3, K_3) \nonumber \\
 & + &  (\epsilon_2, K_3)(\epsilon_3, k_2)
 - { 1\over   \Delta} \left[  2 \, ( K_3^2 K_4^2 - t^2  + \delta )
(\epsilon_2, k_{12}) (\epsilon_3 , k_{12})  \right. \nonumber \\
& + &  (K_3^2 + K_4^2 - s - 2\, t)   \, ( (K_3^2 - K_4^2 + s)
(\epsilon_2 , K_4) (\epsilon_3 , K_4) \nonumber \\
& & \left.  + (K_4^2 - K_3^2 + s) \,
(\epsilon_2 , K_3) (\epsilon_3 , K_3) ) \right] \nonumber \\
& + &
{K_4^2 + t \over K_4^2 - t} \, { (\epsilon_2 , k_1) (\epsilon_3 , k_1)  } +
{K_3^2 + t \over K_3^2 - t} \, { (\epsilon_2 , k_2) (\epsilon_3 , k_2)  } ,
\end{eqnarray}
and %where $k_{12} = k_1 + k_2$
\begin{eqnarray}
 \delta & = & K_3^2 \, K_4^2 - (K_3^2 + K_4^2 )\, t + (s + t)\, t, \\
\Delta  & = & \Delta(k_{12}^2,K_3^2,K_4^2),  \\
\Delta(s_1,s_2,s_3) & = &
s_1^2 + s_2^2 + s_3^2 - 2(s_1s_2+ s_2s_3 + s_3s_1),
\end{eqnarray}
are functions of the external momentum invariants. In particular
$\Delta$ is the Gram determinant of the three-mass triangle
integral arising from the tensor reduction of the two-mass-hard
box integral.

The general case is obtained by changing the reference momentum:
\begin{eqnarray}
 I_4 (\eta \tilde\lambda_1,\epsilon_2,\epsilon_3 )
& = &  { \langle \eta \,  1 \rangle \over \langle   2 \, 1 \rangle }\,
I_4 (\lambda_2\tilde\lambda_1,\epsilon_2,\epsilon_3 ) \nonumber \\
& + & { \langle \eta \,  2 \rangle \over \langle   1
\, 2\rangle } \left[ I_3^{2m} (\epsilon_2,\epsilon_3 ) - I_3^{3m}
(\epsilon_2,\epsilon_3 ) \right] , \label{twomasshard} \\
I_3^{2m} (\epsilon_2,\epsilon_3 )  & = &
 {1\over 2}\, (\epsilon_2, \epsilon_3) +
   {(K^2_3 + t)\over 2 (K^2_3- t)^2 }
  \, (\epsilon_2, k_2) \, (\epsilon_3, k_2)  \nonumber \\
& + & {( (\epsilon_2, k_2)\,(\epsilon_3,K_3)  -(\epsilon_3, k_2)\,
(\epsilon_2, K_4+k_1) ) \over 2(K^2_3-t)} ,
\end{eqnarray}
and $I_3^{3m} (\epsilon_2,\epsilon_3 )$ is the three-mass triangle
integral with external momenta $\{k_{12}, K_3,K_4\}$. There are
two triangle diagrams from this reduction and one is a three-mass
triangle diagram.

\subsection{The box integrals of degree 4 polynomials: the two-mass-easy case}

Now we discuss the computation of the rational parts for degree
4 polynomials.  First we define
\begin{equation}
I_4 (\epsilon_1,\epsilon_2, \epsilon_3, \epsilon_4)   \equiv
 \int { {\rm d}^D p \over i \pi^{D/2}}
\, { (\epsilon_1, p) \, (\epsilon_2 ,  p-K_1) \, (\epsilon_3 ,  p
- K_{12})\, (\epsilon_4, p+K_4)  \over p^2(p-K_1)^2 (p-K_{12})^2
(p+K_4)^2}  ,
\end{equation}
where $K_{12} = K_1 + K_2$. By using Feynman parametrization we
have
\begin{eqnarray}
I_4 (\epsilon_i)  & = & \hat{I}_4[(\epsilon_1,a)(\epsilon_2,a)
(\epsilon_3,a)(\epsilon_4,a)] -
\sum_{i<j}(\epsilon_i,\epsilon_j)\hat{I}_4^{D+2}[
(\epsilon_k,a)(\epsilon_l,a)]  \nonumber \\
&   & \hskip -1cm  +  ( (\epsilon_1,\epsilon_2)
(\epsilon_3,\epsilon_4) + (\epsilon_1,\epsilon_3)
(\epsilon_2,\epsilon_4) + (\epsilon_1,\epsilon_4)
(\epsilon_2,\epsilon_3) ) \, I_{4}^{D+4}[1] .
\end{eqnarray}
We also assume that $\epsilon_{1,3}$ satisfy the physical
conditions in the two-mass-easy box integral and $\epsilon_{1,2}$
satisfy the physical conditions for the 2-mass-hard  box integral.
As we said before this can always be done by expanding all
polarization vectors in terms of the 2  spinors from the 2 massless
momenta.

The direct calculation of the rational part of the two-mass-easy
box integral gives  rather complicated formulas for generic
polarization vectors (even after using the physical condition for
$\epsilon_{1,3}$). By choosing appropriate reference momenta for
$\epsilon_{1,3}$, the result  simplifies greatly. In particular
the reference momentum of $\epsilon_1$ should be $k_3$ and the
reference momentum of $\epsilon_3$ should be $k_1$. In some sense
this is equivalent to the tensor reduction with the two factors
$(\epsilon_1,p)(\epsilon_3,p-k_1-K_2)$. The explicit results are
given as follows:
\begin{eqnarray}
 I_4(\lambda_3\tilde\lambda_1, \epsilon_2,\lambda_1\tilde\lambda_3,
 \epsilon_4)  & = &  -  {1\over 4} \left( {K_2^2 + s \over K_2^2 - s } +
 {K_4^2 + t \over K_4^2 - t }\right) \,(\epsilon_2,k_1)(\epsilon_4,k_1)
 \nonumber \\
&   & \hskip -1.5cm
 - {1\over 4} \left( {K_2^2 + t \over K_2^2 - t
} +  {K_4^2 + s \over K_4^2 - s }\right)
\,(\epsilon_2,k_3)(\epsilon_4,k_3)
-  {5\over 9} \, (k_1,k_3)( \epsilon_2,  \epsilon_4) \nonumber \\
& +  &  {4\over 9}\, \Big( (\epsilon_2,k_1)(\epsilon_4,k_3) +
 (\epsilon_2,k_3)(\epsilon_4,k_1) \Big)   , \\
I_4(\lambda_1\tilde\lambda_3, \epsilon_2,\lambda_1\tilde\lambda_3,
 \epsilon_4)  & = &  {5\over 9}\, \langle 1|\epsilon_2|3] \, \langle1|\epsilon_4|3]
 \nonumber \\
 &  & \hskip -2cm + {\langle1|K_2|3]^2 \over 3}\, \left[
 {(\epsilon_2,k_1)\,(\epsilon_4,k_1) \over (K_2^2 - s) \, (K_4^2 - t)} +
 {(\epsilon_2,k_3)\,(\epsilon_4,k_3) \over (K_2^2 - t) \, (K_4^2 - s)} \right]  .
\end{eqnarray}
Other cases can be either obtained by conjugation or by relabelling
$k_{1,3}$. In fact $I_4(\lambda_1\tilde\lambda_3, \epsilon_2,\lambda_3\tilde\lambda_1,
 \epsilon_4) = I_4(\lambda_3\tilde\lambda_1, \epsilon_2,\lambda_1\tilde\lambda_3,
 \epsilon_4)$ as it is invariant under conjugation.

For the general case, we use the reduction formula by changing the
reference appropriately. For example we have:
\begin{eqnarray}
I_4(\eta_1\tilde\lambda_1,\epsilon_2,\eta_3\tilde\lambda_3,\epsilon_4)
 & = &  {  \langle \eta_1 \, 3\rangle \over \langle 1\, 3\rangle }
\, I_4 (k_1,  \epsilon_2, \epsilon_3, \epsilon_4)   + {\langle
\eta_3 \, 1\rangle   \over \langle 3\, 1\rangle } \, I_4
(\epsilon_1, \epsilon_2, k_3, \epsilon_4)   \nonumber \\
& - &   {\langle \eta_1 \, 1\rangle \, \langle \eta_3 \, 3\rangle \over \langle 1\,
3\rangle^2} \, I_4 (\lambda_3\tilde\lambda_1, \epsilon_2,
\lambda_1\tilde\lambda_3, \epsilon_4)  \nonumber \\
&  + & {\langle \eta_1 \, 3\rangle \, \langle \eta_3 \, 1 \rangle
\over \langle 1\, 3\rangle^2} \,  I_4 (k_1, \epsilon_2, k_3,
\epsilon_4)  .
 \end{eqnarray}
By using the explicit result of the 2-mass triangle integral we have
\begin{eqnarray}
& &  \hskip - 2cm I_4(\epsilon_1,\epsilon_2,k_3,\epsilon_4)
  =   { K_2^2 + s \over 6 ( K_2^2 - s)^2} \,
(\epsilon_1, K_2)(\epsilon_2,k_1)(\epsilon_4,k_1) \nonumber \\
& + & { K_4^2 + t \over 6 ( K_4^2 - t)^2} \,
 (\epsilon_1, K_4)(\epsilon_2,k_1)(\epsilon_4,k_1)
\nonumber \\
& +  & {1\over 12}\,  \left( {K_2^2 + s \over  K_2^2 - s  }
 +{ K_4^2 + t \over  K_4^2 - t  } \right)
 \, ((\epsilon_1, \epsilon_2) (\epsilon_4,k_1) +
 (\epsilon_1, \epsilon_4)(\epsilon_2,k_1) \nonumber \\
& + &
{ (\epsilon_1, K_2)  \over 6 ( K_2^2 - s) } \,(\epsilon_2,k_1)(\epsilon_4,k_3)
 +{ (\epsilon_1, K_4)  \over 6 ( K_4^2 - t) } \,
 (\epsilon_2,k_3)(\epsilon_4,k_1)  \nonumber \\
 & + &  \left[ {(\epsilon_1,K_2)\over 6\,(K_2^2 - s)}  -
  {(\epsilon_1,K_4)\over 6\,(K_4^2 - t)}
 \right] \, ( (\epsilon_2,k_1)\,(\epsilon_4,K_4) -
 (\epsilon_2,K_2)\,(\epsilon_4,k_1) ) \nonumber \\
  & + & {1\over 9} ( (\epsilon_1,\epsilon_2)(\epsilon_4,k_3) +
(\epsilon_1,\epsilon_4)(\epsilon_2,k_3) +
(\epsilon_2,\epsilon_4)(\epsilon_1,k_3) )\nonumber \\
& + & {1\over2} \, (\epsilon_2,k_1)\,(\epsilon_4,K_4)\, \left[
{(\epsilon_1, K_4)\over K_4^2 - t }  -  {(\epsilon_1, K_2)\over
K_2^2 - s } \right] .
 \end{eqnarray}
Except for the last term, this formula is invariant under the
interchange $2 \leftrightarrow 4$ ($ s\leftrightarrow t$). By
setting $\epsilon_1 = k_1$ we get
\begin{equation}
I_4(k_1,\epsilon_2,k_3,\epsilon_4) = {1\over 18}( 2 (k_1,k_3)(\epsilon_2,\epsilon_4) - (
(\epsilon_2,k_1)(\epsilon_2,k_3) + (\epsilon_2,k_3)(\epsilon_2,k_1) ) ) .
\end{equation}
This agrees with the result by direct computation by first doing the tensor reduction
and then using the result for bubble integrals. These formulas are quite useful for
obtaining compact analytic formulas for QCD amplitudes.

For easy reference we also give here the relevant formulas in terms of
Feynman parameters:
\begin{eqnarray}
& & \hskip -1cm \hat{I}_4
[(\tilde\epsilon_1,a)(\epsilon_2,a)(\tilde\epsilon_3,a)(\epsilon_4,a)]
\nonumber \\
& = & {(\tilde\epsilon_1, K_2)\,(\tilde\epsilon_3 ,
K_2) \over 3} \, \left( {(\epsilon_2, k_1) \, (\epsilon_4, k_1)
\over (K_2^2 - s)(K_4^2 - t)} + {(\epsilon_2, k_3) \, (\epsilon_4,
k_3)
\over (K_2^2 - t)(K_4^2 - s)} \right) \nonumber \\
& & \hskip -1cm \hat{I}_4^{D+2}[(\epsilon_2,a)(\epsilon_4,a)] = {1\over
6(K_2^2 + K_4^2 - s- t) }\Big[ (\epsilon_2, k_1)\, (\epsilon_4 ,
k_3) +
(\epsilon_2, k_3)\, (\epsilon_4 , k_1) \nonumber \\
&  - &  (K_2^2 \, K_4^2 - s\, t)\Big( { (\epsilon_2, k_1)\,
(\epsilon_4 , k_1) \over (K_2^2 - s)(K_4^2 - t)} +{ (\epsilon_2,
k_3)\, (\epsilon_4 , k_3) \over (K_2^2 - t)(K_4^2 - s)} \Big)\Big]
.
\end{eqnarray}
All other $\hat{I}_4^{D+2}[(\epsilon_i,a)(\tilde\epsilon_j,a)]$'s are
identically zero. In the above formulas $\tilde\epsilon_{1,3}$
satisfy 2 conditions: $(\tilde\epsilon_i,k_{1,3})=0$, i.e. the
reference momentum of $\tilde\epsilon_1$ is $k_3$ and  the
reference momentum of $\tilde\epsilon_3$ is $k_1$.  We also note
when $(\tilde\epsilon_1,\tilde\epsilon_3)$ is not equal to 0, it
cancels with the factor $(K_2^2 + K_4^2 - s - t) = (k_1,k_3) = -
(\tilde\epsilon_1, \tilde\epsilon_3)$.

\subsection{The box integrals of degree 4 polynomials: the two-mass-hard case}

As before, the direct computation of the rational part of the two-mass-hard
 box gives very complicated algebraic expressions. In
fact it gives the  most complicated formula up to now. Changing the
reference momenta simplifies a little bit, but the resulting
formula is still not intelligible. The main complication comes from
the presence of the 3-mass triangle integrals. To organize the
final result, we proceed to do tensor reduction one more time. To begin
with let us  define the integral we want to compute:
\begin{eqnarray}
I_4^{2mh}(\epsilon_1,\epsilon_2,\epsilon_3,\epsilon_4;c_3,c_4)  & & \nonumber \\
& & \hskip -4cm \equiv I_4[ (\epsilon_1,p)
(\epsilon_2,p-k_1)((\epsilon_3, p+K_4)+c_3)((\epsilon_4, p+K_4)+c_4)] \nonumber \\
& & \hskip -4cm = \int{ {\rm d}^Dp \over i \pi^{D/2}} \, {
(\epsilon_1,p) (\epsilon_2,p-k_1)((\epsilon_3,
p+K_4)+c_3)((\epsilon_4, p+K_4)+c_4) \over p^2\, (p-k_1)^2\,
(p-k_{12})^2\, (p+K_4)^2 } . \nonumber \\
\end{eqnarray}
The external momenta are $k_1$, $k_2$, $K_3$ and $K_4$ as shown in
Fig.~\ref{Eight} by setting $K_1=k_1$ and $K_2 =k_2$. $k_{1,2}$
are the two massless external legs and $K_{3,4}$ are the two
massive external legs. We set $t=(k_2 +K_3)^2=(K_4+k_1)^2$. We
also require that $\epsilon_{1,2}$ satisfy the physical condition,
i.e. $(\epsilon_1,k_1)=0$ and $(\epsilon_2,k_2)=0$. The formula
for generic $\epsilon_{1,2}$ is beyond the scope of this paper.
($\epsilon_{3,4}$ are arbitrary 4-dimensional polarization
vectors.)

There are 4 possible cases for $\epsilon_{1,2}$. In the same helicity
cases, the box integrals can be easily reduced to triangle integrals
by using the reduction formulas eq. (\ref{eqreduction}) and (\ref{eqreductiona}) in Sect. 3.
So we will only consider the difficult cases
where $\epsilon_{1,2}$ have different helicities. To be definite
we set  $\epsilon_1 = \lambda_1 \tilde\eta_1 $ and $\epsilon_2 =
\eta_2\tilde\lambda_2$. The opposite case can be obtained from
this one simply by conjugation. We will give the explicit formula
for this case at the end of this subsection.

To do tensor reduction for the factors associated with the two
massless external legs, we have:
\begin{eqnarray}
T^{-+} & = &  (\epsilon_1,p)\,(\epsilon_2,p-k_1) =
{\langle\eta_2|(p-k_1)\,k_2\,K_3\,k_1\,
(p-k_1)|\tilde\eta_1\rangle\over \langle  2| K_3 |1]}.
\end{eqnarray}
The middle factor in the above can be decomposed by moving the
first factor of $(p-k_1)$ towards the second $(p-k_1)$. Explicitly
we have:
\begin{eqnarray}
& & \hskip -2cm (p-k_1)\,k_2\,K_3\,k_1\, (p-k_1) =  T_1 + T_2, \\
 T_1 & = &  I^{(1)} \, (p-k_1)k_2K_3 + I^{(2)}\,(K_3k_1-k_2K_3) p + I^{(3)}K_4k_1 p,
\\
T_2 & = &  p_{-2\epsilon}^2 k_2K_3k_1 + (I^{(4)}- t) k_2k_1 p  \nonumber \\
& + &  I^{(1)}( - I^{(2)}K_3 + I^{(3)}(k_2+K_3) - (I^{(4)} - t)
k_2) .
\end{eqnarray}
Omitting the factor $\langle  2| K_3 |1]$, we now compute the
various terms. We have:
\begin{eqnarray}
\langle\eta_2| T_2|\tilde\eta_1\rangle
& = &  p_{-2\epsilon}^2 \langle\eta_2| k_2K_3k_1|\tilde\eta_1\rangle \nonumber \\
& + & (I^{(4)}- t) (
 \langle\eta_2\, 2\rangle \, [\tilde\eta_1\, 1] (\lambda_1\tilde\lambda_2,p) -
 I^{(2)} \, \langle\eta_2|k_2| \tilde\eta_1 \rangle )
 \nonumber \\
& +  & I^{(1)}\, ( I^{(3)}\,  \langle\eta_2|(k_2+K_3)|
\tilde\eta_1 \rangle  -
 I^{(2)} \,  \langle\eta_2|K_3| \tilde\eta_1 \rangle ) .
\end{eqnarray}
This gives the following rational terms:
\begin{eqnarray}
A_2 & = & - \frac{1}{6} \, \langle\eta_2|
k_2K_3k_1|\tilde\eta_1\rangle \, (\epsilon_3,\epsilon_4) -
t\,\langle\eta_2\,2\rangle \, [\tilde\eta_1\,1] \,
I_4^{2mh}(\lambda_1\tilde\lambda_2,\epsilon_3,\epsilon_4)
\nonumber \\
& + &  t \, \langle\eta_2|k_2|\tilde\eta_1 ] \,
I_3^{3m}(\epsilon_3,\epsilon_4) +  \langle\eta_2\,2\rangle \,
[\tilde\eta_1\,1] \left( {1\over 2}
(\langle1|\epsilon_3|2] \, c_4 + \langle1|\epsilon_4|2] \, c_3) \right. \nonumber \\
& & \left. + {1\over 18} (\langle1|\epsilon_3|2] \, \epsilon_4 +
\langle1|\epsilon_4|2] \,
\epsilon_3, 7\,k_1+2\,k_2 + 9\,K_4)\right) \nonumber \\
& + & {1\over18}\, \langle  \eta_2\,2\rangle \,[\tilde\eta_1\,2]
\, ( (\epsilon_3,k_{12})\,(\epsilon_4,k_{12}) - 2 \, s_{12} \,
(\epsilon_3,\epsilon_4) ) \nonumber \\
  & + & \frac{1}{18}\, \langle\eta_2|(k_2+K_3)|\tilde\eta_1] (
(\epsilon_3, k_2+K_3)(\epsilon_4,
 k_2+ K_3) - 2 \, t\, (\epsilon_3, \epsilon_4) ) \nonumber \\
& - & \frac{1}{18}\, \langle\eta_2|K_3|\tilde\eta_1] \, (
(\epsilon_3, K_3)(\epsilon_4, K_3) - 2
K_3^2(\epsilon_3,\epsilon_4))
\end{eqnarray}
The explicit formulas for the other 3 terms in $T_1$ are not quite
illuminating and we refrain from writing the explicit results here. We
can write the result in terms of the rational parts for the three-mass
and two-mass triangle integrals arising from the tensor reduction.
We have:
\begin{eqnarray}
A_{1 } & = & \langle 2|K_3|\tilde\eta_1\rangle (
I_3^{2m}(\eta_2\tilde\lambda_2,
\epsilon_3,\epsilon_4) \nonumber \\
& + & I_3^{2m} (\eta_2\tilde\lambda_2,
(c_3-(\epsilon_3,K_3))\epsilon_4 + (c_4
+ (\epsilon_4,K_4+k_1)) \epsilon_3) ) \nonumber \\
& + &  I_3^{3m}(v, \epsilon_3,\epsilon_4) +  I_3^{3m}(v,
(c_3-(\epsilon_3,K_3))\epsilon_4 + c_4\epsilon_3 ), \\
 & + &  \langle\eta_2|K_4|1]\, ( \tilde I_3^{2m}(\lambda_1\tilde\eta_1,\epsilon_3,
\epsilon_4)\nonumber \\
& + &\tilde
I_3^{2m}(\lambda_1\tilde\eta_1,(c_3-(\epsilon_3,k_2+K_3))\epsilon_4+
(c_4+(\epsilon_4,K_4))\epsilon_3 ),
\end{eqnarray}
where
\begin{equation}
v =  \langle\eta_2|K_3|1] \lambda_1\tilde\eta_1 +
\langle\eta_2|K_3|2] \lambda_2\tilde\eta_1
-(k_2,K_3)\eta_2\tilde\eta_1 .
\end{equation}
Combining the above results together we have:
\begin{equation}
I_4^{2mh}(\lambda_1\tilde\eta_1,\eta_2\tilde\lambda_2,\epsilon_3,\epsilon_4;
c_3,c_4) = {A_1 + A_2 \over \langle  2| K_3 |1]} .
\end{equation}
For the opposite helicity case the  formula is:
\begin{eqnarray}
& & \hskip -1cm
I_4^{2mh}(\eta_1\tilde\lambda_1,\lambda_2\tilde\eta_2,\epsilon_3,\epsilon_4;
c_3,c_4) = {1
 \over \langle  1| K_3 |2]} \left[
 - \frac{1}{6} \, \langle\eta_1| k_1K_3k_2|\tilde\eta_2\rangle
\, (\epsilon_3,\epsilon_4) \right.
\nonumber \\
& - & t\,\langle\eta_1\,1\rangle \, [\tilde\eta_2\,2] \,
I_4^{2mh}(\lambda_2\tilde\lambda_1,\epsilon_3,\epsilon_4)
+   t \, \langle\eta_1|k_2|\tilde\eta_2 ] \, I_3^{3m}(\epsilon_3,\epsilon_4) \nonumber \\
& + & \langle\eta_1\,1\rangle \, [\tilde\eta_2\,2] \left( {1\over
2}
(\langle2|\epsilon_3|1] \, c_4 + \langle2|\epsilon_4|1] \, c_3) \right. \nonumber \\
& & \left. + {1\over 18} (\langle2|\epsilon_3|1] \, \epsilon_4 +
\langle2|\epsilon_4|1] \,
\epsilon_3, 7k_1+2k_2 + 9K_4)\right) \nonumber \\
& + & {1\over18}\, \langle  \eta_1\,2\rangle \,[\tilde\eta_2\,2]
\, ( (\epsilon_3,k_{12})\,(\epsilon_4,k_{12}) - 2 \, s_{12} \,
(\epsilon_3,\epsilon_4) ) \nonumber \\
& + & \frac{1}{18}\, \langle\eta_1|(k_2+K_3)|\tilde\eta_2] (
(\epsilon_3, k_2+K_3)(\epsilon_4,
 k_2+ K_3) - 2 \, t\, (\epsilon_3, \epsilon_4) ) \nonumber \\
& - & \frac{1}{18}\, \langle\eta_1|K_3|\tilde\eta_2] \, (
(\epsilon_3, K_3)(\epsilon_4, K_3) - 2
K_3^2(\epsilon_3,\epsilon_4))  \nonumber \\
& + & \langle\eta_1|K_3|2] (
I_3^{2m}(\lambda_2\tilde\eta_2,
\epsilon_3,\epsilon_4) \nonumber \\
&   & +  I_3^{2m} (\lambda_2\tilde\eta_2,
(c_3-(\epsilon_3,K_3))\epsilon_4 + (c_4
+ (\epsilon_4,K_4+k_1)) \epsilon_3) ) \nonumber \\
& + &  I_3^{3m}(\tilde v, \epsilon_3,\epsilon_4) +
I_3^{3m}(\tilde v,
(c_3-(\epsilon_3,K_3))\epsilon_4 + c_4\epsilon_3 ), \\
 & + &  \langle1|K_4|\tilde\eta_2\rangle\, ( \tilde I_3^{2m}(\eta_1\tilde\lambda_1,\epsilon_3,
\epsilon_4)\nonumber \\
&   & + \left. \tilde
I_3^{2m}(\eta_1\tilde\lambda_1,(c_3-(\epsilon_3,k_2+K_3))\epsilon_4+
(c_4+(\epsilon_4,K_4))\epsilon_3 ) \right] ,
\end{eqnarray}
where
\begin{equation}
\tilde v =\langle1|K_3|\tilde\eta_2\rangle \,\eta_1\tilde\lambda_1
+ \langle2|K_3|\tilde\eta_2\rangle \, \eta_1\tilde\lambda_2
-(k_2,K_3)\eta_1\tilde\eta_2 .
\end{equation}

\section{Extra terms for box and triangle tensor reduction}

\subsection{Extra terms for box tensor reduction}

In order to get a comparatively compact analytic formula for the
two-mass-hard box integral, it is necessary to do tensor reduction
because a direct computation of the rational part by the recursive
method gives a very complicated formula. It would be a better idea
to organize it into lower degree and lower point integrals. A naive
tensor reduction directly in $D=4$ would give an incorrect result
because the box integral is ultra-violet divergent for a degree 4
polynomial of momenta in the numerator. It turns out that the
difference between  the $D=4$ tensor reduction and the correct
tensor reduction is just a rational function. In this subsection
we will compute this extra rational function explicitly.

For a degree 2 polynomial $g(\epsilon,k,p)$ in the internal momentum $p$,
the general form of the $D=4$ tensor reduction we used is as follows:
\begin{equation}
g(\epsilon, k, p)  = \tilde g(\epsilon,k, p)
- p^2\, f(\epsilon,k) . \label{extra}
\end{equation}
We assume that all polarization vectors $\epsilon$ and momenta $k$
are 4-dimensional and the above relation is derived by assuming
$p$ is also a 4-dimensional momentum. We also assume that $g$ and
$\tilde g$ depend on the momentum $p$ through scalar products
$(\epsilon,p)$ and/or $(k,p)$. Because we use FDH regularization
\cite{FDH}, $p$ is actually promoted to be in $D=4-2\epsilon$
dimensions in our later calculations of the rational part. This
affects only the last term in eq.~(\ref{extra}).   For arbitrary
dimensional internal momentum $p$, the above formula is still
valid if we make the substitution:
\begin{equation}
p^2 \rightarrow p^2 - p^{2}_{D-4}.
\end{equation}
That is, the extra dimensional part ($ D-4  = - 2\epsilon$) of the
momentum $p$ must be subtracted from $p^2$ which actually stands
for  the scalar product in $D$ dimensions in the subsequent
computation in the four-dimensional helicity regularization scheme.
Because pentagon and hexagon diagrams are ultra-violet convergent
we can safely discard this term in the tensor reduction for 5- or
6-point diagrams.  This term does give a non-vanishing
contribution to the rational parts for box and triangle integrals.

For box integral we have
\begin{eqnarray}
A_{\rm box} & = & I_4^{D=4-2\epsilon}[ p^2_{-2\epsilon}
\, (\epsilon_3,p) \, (\epsilon_4,p) ]
\nonumber\\
& = & \int_0^1 {\rm d}^4 a_i \delta(1 - \sum_i a_i) \,
\int_0^\infty{\rm d}T \, T^3
\nonumber \\
& & \times
\int { {\rm d}^D p \over i \pi^{D/2} } \, p^2_{-2\epsilon} \,
(\epsilon_3, p) \,
(\epsilon_4, p) \, {\rm e}^{ p^2 T - T \, a\cdot S\cdot a},
\end{eqnarray}
by transforming it into a Feynman parameter integral and omitting terms
which are vanishing in the limit $\epsilon \to 0$.

The integration over the momentum $p$ can be done easily by splitting it into a $D=4$
part and a  $(-2\epsilon)$-dimensional part. We have then
\begin{equation}
A_{\rm box} =  \epsilon\, ( -  (\epsilon_3, \epsilon_4) ) \,
 I_4^{D+4}[1] =
- { 1\over 6} \, (\epsilon_3, \epsilon_4)  + O(\epsilon).
\end{equation}
By using this result the correct formula for computing the
rational part of the box tensor integral is:
\begin{eqnarray}
  I_4[ g(\epsilon, k, p)\,
( (\epsilon_3,p) +c_3)\, ((\epsilon_4, p) + c_4)) ]
 & = &
 I_4[ (\tilde g(\epsilon,k,p) - p^2\, f(\epsilon,k) ) \,
\nonumber \\
& & \hskip -6cm \times ( (\epsilon_3,p) +c_3)\, ((\epsilon_3, p) + c_4)) ] -
{ 1\over 6} \, f(\epsilon, k) \, (\epsilon_3,
\epsilon_4) .
\end{eqnarray}

\subsection{Extra terms in triangle tensor reduction}
For triangle integrals, using the  same reduction formula as given
in eq.~(\ref{extra}) also gives incorrect results. The extra terms
are given by the following formula ($p$ is the internal momentum
between the external legs $K_1$ and $K_2$):
\begin{eqnarray}
I_3[ g(\epsilon, K, p)\,
( (\epsilon_3,p) + c_3)  ]
 & = & I_3[ ( \tilde g(\epsilon, K, p) - p^2\, f(\epsilon,k) ) \,
\nonumber \\
& & \hskip -4cm \times ( (\epsilon_3,p) +c_3)  ] - f(\epsilon,k)\left[
{1\over 2} \, c_3  + {1\over 6} \, (\epsilon_3, K_1-K_3) \right] .
\end{eqnarray}
We note that the rational part  $I_3[g(\epsilon, K, p)\,
( (\epsilon_3,p) + c_3)  ]$ is defined by the Feynman integral without
the usual minus sign, just as we did before in eq.~(\ref{ithreemass}).

\section{Conclusion}

The calculation of multi-leg one-loop amplitudes in QCD is  highly motivated
through the need of precise predictions for
multi-particle scattering at TeV colliders like the
Tevatron and the LHC. Whereas unitarity based methods
are very successful to provide the  cut-constructible
parts of one-loop amplitudes, the evaluation of
rational terms which are not defined by the cuts alone,
is a much harder question. Recently  the bootstrap recursive proposal of Bern,
Dixon and Kosower \cite{BDKBoot} has been made for their evaluation. (It was used
to do real QCD calcuations \cite{BDKBoot,BDKE}.)
In this paper we have developed a general formalism targeting only the computation of
the rational terms by using Feynman diagrammatic
methods.

In comparing with the previous use of diagrammatic
methods in the computation of multi-leg one-loop amplitudes, the computation of the
rational
terms  only is greatly simplified by using the BDDK theorem \cite{BDDK}. By using this
theorem and
a thorough analysis of the recursive relations we derived recursive relations also
for the rational terms.
Quite explicit results are given for all bubble and triangle integrals, and box
integrals up to two-mass cases.
These results are all the ingredients for computing up to the 6-gluon one-loop
amplitudes in QCD.
We will compute the (rational parts of the) 5-gluon and 6-gluon amplitudes in
\cite{xyzii} and \cite{xyziii}
respectively.

By using the  recursive relations it is straightforward to derive the rational terms
of the 3-mass and 4-mass
(tensor) box integrals. The formulas obtained are quite complicated. It may only be
useful for implementing numerically.

The method developed in this paper
can be extended to the cases with massive internal particles
and internal (massive) quarks.
Our method complements the twistor-inspired approach quite nicely.  We envisage
that an automatic implementation of both approaches could push the present limit of
6-gluon to about 8-gluon or higher.

\section*{Acknowledgments}

We would like to thank Carola F. Berger, Zvi Bern, Lance J. Dixon,
Darren Forde and David A. Kosower for sending us their results
prior to publication \cite{BDKE}, and for discussions, reading the
paper, comments and assistance in comparing our six-gluon results
with theirs. We also thank them for checking the NMHV results by
using their factorization check program.  CJZ would like to thank
J. -P. Ma for constant encouragements, helpful discussions and
careful reading of the paper. His (financial) support (to buy a
computer which was still in use today) actually goes back to the
much earlier difficult times when I did not have enough grants from
other sources. What is more important is that there are no strings
attached to his support and it is up to the last author to explore
what he wants to. CJZ would also like to thank R. Iengo for
encouragements and his interests in this work, helpful discussions
and comments; to Z.~Chang, B.~Feng, E.~Gava, H.~-Y.~Guo,
K.~S.~Narain, K.~Wu, Y.~-S.~Wu, Z.~Xu and Z.~-X.~Zhang for
discussions and comments; to Prof.~X.~-Q.~Li and the hospitality
at Nankai University where we can have good food; and finally to
Prof.~S.~Randjbar-Daemi and the hospitality at Abdus Salam
International Center for Theoretical Physics, Trieste, Italy. This
work is supported in part by funds from the National Natural
Science Foundation of China with grant number 10475104 and
10525522.

\end{document}